\documentclass[12pt,aps,english]{revtex4}
\usepackage{graphicx}
\usepackage{bm}
\usepackage{dcolumn}
\usepackage[usenames, dvipsnames]{color}
\usepackage{epsfig}
\begin{document}

\title{Reconfiguration Dynamics in folded and intrinsically disordered protein with internal friction: Effect of solvent quality and denaturant}
\author{Nairhita Samanta and Rajarshi Chakrabarti*}
\affiliation{Department of Chemistry, Indian Institute of Technology Bombay, Mumbai, Powai 400076, E-mail: rajarshi@chem.iitb.ac.in}
\date{\today}

\begin{abstract}

We consider a phantom chain model of polymer with internal friction in a harmonic confinement and extend it to take care of effects of solvent quality following a mean field approach where an exponent $\nu$ is introduced. The model termed as ``Solvent Dependent Compacted Rouse with Internal Friction (SDCRIF)" is then used to calculate the reconfiguration time of a chain that relates to recent F\"{o}rster resonance energy transfer (FRET) studies on folded and intrinsically disordered proteins (IDPs) and can account for the effects of solvent quality as well as the denaturant concentration on the reconfiguration dynamics. Following an ansatz that relates the strength of the harmonic confinement ($k_c$) with the internal friction of the chain ($\xi_{int}$), SDCRIF can convincingly reproduce the experimental data and explain how the denaturant can change the time scale for the internal friction. It can also predict near zero internal friction in case of IDPs. In addition, our calculations show that the looping time as well as the reconfiguration time scales with the chain length $N$ as $\sim N^\alpha$, where $\alpha$ depends weakly on the internal friction but has rather stronger dependence on the solvent quality. In absence of any internal friction, $\alpha=2\nu+1$ and it goes down in presence of internal friction, but looping slows down in general. On the contrary, poorer the solvent, faster the chain reconfigures and forms loop, even though one expects high internal friction in the collapsed state. However, if the internal friction is too high then the looping and reconfiguration dynamics become slow even in poor solvent.

\end{abstract}

\maketitle
\section{Introduction}

Among the polymer rheologists the notion of internal friction associated with a single polymer chain is more than twenty five years old \cite{rabin1990, schieber, schieber2, manke1985}. Surprisingly it is only very recently that the topic has gained attention in the chemical and biophysics community.  Although some earlier works in chemical physics community did point out the importance of internal friction in single chain dynamics \cite{thirumalaijcp2000, wolynes2001} but did not receive the attention it should have. However recent  experiments on looping  and folding dynamics \cite{hagen2004, hagen2005, hagen2007, Beddard2015, schuler2012, bhuyan2013} in proteins have indicated the non-negligible role of internal friction. In this context it is worth mentioning that loop formation between any two parts of a bio-polymer \cite{wilemski1974, szabo, ghosh2002, ghosh2003, portman, sokolov2003, toan, sebastianjcp, snigdha2014, metzler2015, Cherstvy2011, Metzlermacrolet2015} is supposedly the primary step of protein folding, DNA cyclization and since internal friction affects looping it is obvious that measurements of folding rates in proteins would predict the importance of solvent independent internal friction as well \cite{eatonpnas2008}.  Subsequently not only these recent experiments \cite{schuler2012, schulernaturecomm2012, bhuyan2013} on polypeptides and proteins showed internal friction to play a pivotal role in the dynamics but also motivated theoretical chemical physicists to come up with statistical mechanical models for single polymer chain with the inclusion of internal friction \cite{makarov2010, makarov2013, chakrabarti2013, chakrabarti2014, chakrabarti2015} and apply these models to investigate the loop formation dynamics.  Other than model build up there have been attempts to elucidate the origin of internal friction in proteins based on computer simulation studies \cite{netzepje, netzjacs, netzbiophysj2013, best2014, makarov2014, best2015, netzjpcb2015}.  A careful literature survey would reveal that it was de Gennes \cite{degennes}, who introduced the concept of internal viscosity at the single chain level. Rabin and  \"{O}ttinger proposed an expression for the relaxation time, $\tau_{rel}$ associated with internal viscosity following an idea of de Gennes \cite{degennes}, which is $\tau_{rel}=R^3/k_BT(\eta_s+\eta_i)$ where, $R=aN^{\nu}$ and $a$, $N$ are the monomer size and chain the length respectively, $\nu$ is the Flory exponent \cite{floryppc, florysmcm, rubinstein}. Therefore in the limit solvent viscosity $\eta_s\rightarrow 0$,  it has a non-zero intercept proportional to the internal viscosity $\eta_i$. This is what exactly seen in recent experiments where the plot of reconfiguration time vs solvent viscosity has a finite intercept equal to the time scale for the internal friction. To the best of our knowledge so far all the theoretical attempts on loop formation in a single chain with internal friction have been restricted to $\theta$ solvent, with $\nu=1/2$, when the chain behaves ideally. But the experimental conditions remain close to a good solvent rather than a $\theta$ solvent. There have been few theoretical studies to elucidate the effect of solvent quality on loop formation in single polymer chain \cite{lee2003, cherayiljcp2004, lee2008} and unfortunately there exists almost no theoretical study to analyze the combined effect of solvent quality and internal friction on the ring closure dynamics in polymer chains other than the very recent simulation by Yu and Luo \cite{luo2015}. Apart from the solvent quality, denaturant concentration does play an important role in ring closure dynamics \cite{chakrabarti2015, eatonbiophysj2006} of proteins as it profoundly affects the compactness of the protein and routinely used in experiments. Very recently Samanta and Chakrabarti \cite{chakrabarti2015}  used a compacted Rouse chain model with internal friction to infer the role of denaturant on ring closure dynamics. But the model works only in the $\theta$ solvent condition where a phantom Rouse chain description works.  Experiments have been performed with proteins away from the $\theta$ conditions where excluded volume interactions along with the internal friction play important roles. To take care of excluded volume interactions one has to go beyond phantom chain description but then the many body nature of the problem does not allow an analytical solution. One possibility would be to perform computer simulation as is recently done by Yu and Luo \cite{luo2015}. Other possibility would be to work with a polymer chain where the excluded volume interactions are taken care of at the mean field level. We take the second route and propose a very general model that takes care of solvent quality as well as denaturant in addition to internal friction.  We call it ``Solvent Dependent Compacted Rouse with Internal Friction (SDCRIF)". To take care of solvent effect we closely follow the work of Panja and Barkema \cite{barkema2009} where an approximate analytical expression for the end to end vector correlation function for a flexible chain in an arbitrary solvent was proposed based on a series of computer simulations. The expression carries a parameter $\nu$ similar in the spirit of Flory exponent \cite{floryppc, florysmcm, rubinstein, plaxco2004, hofmann2012} which takes care of solvent quality.  A value of $\nu=1/2$ corresponds to a $\theta$ solvent and in that case the correlation function is exact and reduces to the text book expression for the ideal chain \cite{doibook, kawakatsubook}, on the other hand $\nu=3/5$ ($0.588$ more precisely) \cite{barkema2009} corresponds to a self avoiding flexible chain (good solvent) as is the case with real polymers. Importantly the same expression can be used for a range of values of $\nu$ corresponding to different solvent qualities.  Very recently such a mean field Flory exponent based model has been used to describe ring polymer dynamics \cite{rubinstein2015}. Next is the inclusion of internal friction which is done similarly as in case of a phantom polymer chain \cite{khatri2007}. Further to take care of the denaturant which controls the compactness of the protein a confining harmonic potential with force constant $k_c$ is used as is done in a very recent study by the authors \cite{chakrabarti2015} and also in the context of diffusing polymers in microconfinements \cite{holcman2013} or in case of bubble formation in double stranded DNA \cite{chakrabarti}. So the novelty of SDCRIF remains in its applicability for a range of solvent quality, denaturant concentration and in addition it also takes care of internal friction ($\xi_{int}$) when required. SDCRIF reduces to Rouse model in the limit $\nu=1/2, k_c=0, \xi_{int}=0$ and to `` Compacted Rouse with Internal Friction (CRIF)"  \cite{chakrabarti2014} in the limit $k_c\neq0, \xi_{int}\neq0, \nu=1/2$.  The looping dynamics is studied within Wilemski Fixman (WF) framework \cite{wilemski1974, doi} with SDCRIF, assuming the polymer chain to be Gaussian. It is worth mentioning that WF formalism has extensively been used to calculate the average ring closure or looping time in presence of hydrodynamic interactions by Chakrabarti \cite{chakrabartiphysica1}  and to elucidate the effect of viscoelastic solvent \cite{chakrabartiphysica2, bhattacharyya} and even applied to ring closing of a semiflexible chain \cite{santo}.

The arrangement of the paper is as follows. The model SDCRIF is introduced in section $\bf{II}$. Section $\bf{III}$ deals with the method used to calculate the reconfiguration and ring closure time. Results and discussions are presented in section $\bf{IV}$ and section $\bf{V}$ concludes the paper.

\begin{figure}
\centering
 \includegraphics[width=0.8\textwidth]{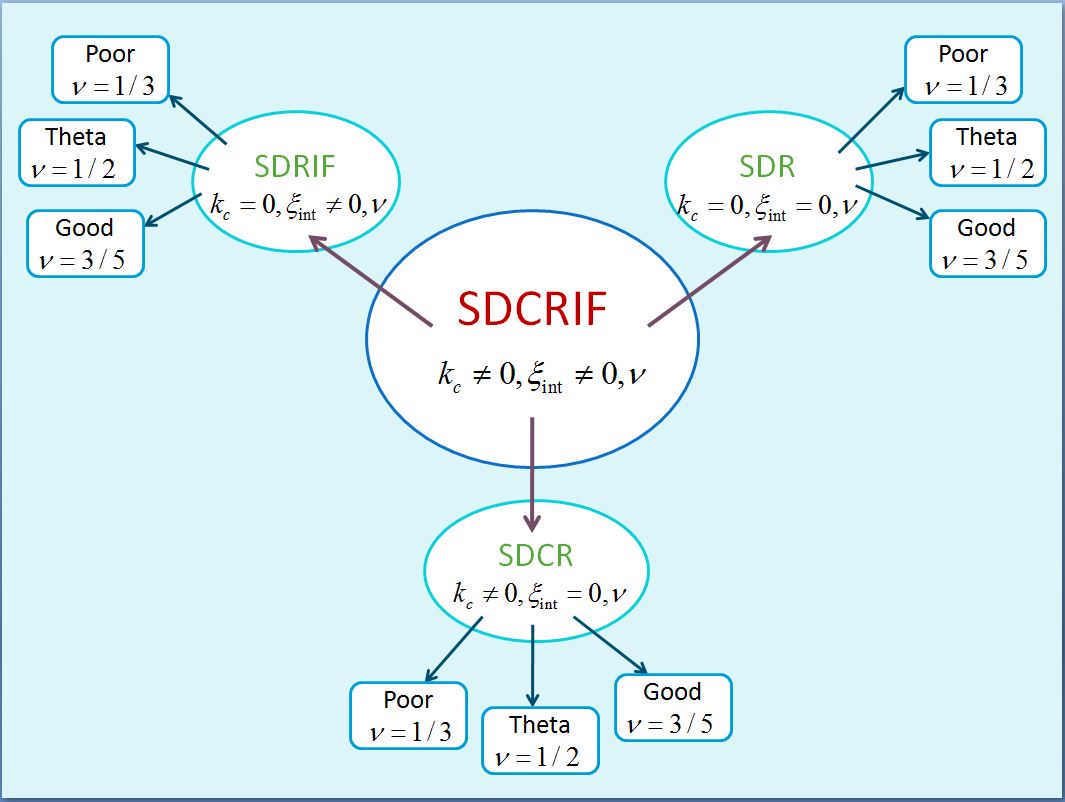}
 \caption{Solvent Dependent Compacted Rouse with Internal Friction (SDCRIF)}
 \label{fig:a}
 \end{figure}

\section{Solvent quality dependent compacted Rouse with Internal friction (SDCRIF)}

In the Rouse model, the polymer chain is treated as a phantom chain  \cite{doibook, kawakatsubook}, where the hydrodynamics interactions and the excluded volume effects are not present.  If $R_n(t)$ denotes the position of the $n^{th}$ monomer at time $t$ of such a chain, made of $(N+1)$ monomers, $n$ varying from $0$ to $N$. The equation describing the dynamics of the chain is the following

\begin{equation}
\xi \frac{\partial{R_{n}(t)}}{\partial{t}}=k\frac{\partial^2{R_{n}(t)}}{\partial{n^2}}+f(n,t)
\label{eq:rouse-model}
\end{equation}

\noindent Where $\xi$ denotes the friction coefficient and $k=\frac{3 k_B T}{b^2}$ is the spring constant where $b$ is the Kuhn length. $f(n,t)$ is the random force with moments

\begin{equation}
   \left<f(n,t)\right>=0,
   \left<f_{\alpha}(n,t_1)f_{\beta}(m,t_2)\right>=2 \xi k_B T \delta_{\alpha\beta}\delta(n-m)\delta(t_1-t_2)
\label{eq:random-force}
\end{equation}

The above equation is solved by decomposing it into normal modes as follows $R_{n}(t)={X_0} + 2 \sum\limits_{p=1}^\infty X_p(t)cos(\frac{p \pi n}{N})$. In the normal mode description the above force balance equation transforms to the following

\begin{equation}
\xi_p \frac{d{X_{p}(t)}}{d{t}}=-k_p X_{p}(t)+f_p(t)
\label{eq:rouse-mode}
\end{equation}

The time correlation function of the normal modes is

\begin{equation}
\left<X_{p\alpha}(0)X_{q\beta}(t)\right>=\frac{k_BT}{k_p}\delta_{pq}\delta_{\alpha\beta}\exp\left(-t/\tau_p\right)
\label{eq:xpcorr}
\end{equation}

\noindent The above expression is for a phantom polymer chain, a chain in $\theta$ solvent. In reality $\theta$ condition is rarely achieved and the chain behaves as in a good solvent where the chain swells due to the excluded volume interactions. For example Buscaglia et. al showed \cite{eatonbiophysj2006} that to a 11 residue polypeptide chain, 6M GdmCl and 8 M urea are good solvents but aqueous buffer is very close to a $\theta$ solvent. It is also not uncommon to encounter a situation representing more of a bad solvent where the chain has a collapsed conformation. To take care of  solvent quality we introduce an exponent $\nu$ similar in the spirit of Flory exponent \cite{doibook} in the above expression for the time correlation function of the normal modes.  This inclusion closely follows the work of Panja and Barkema \cite{barkema2009}. Apart from the solvent quality the presence of denaturant also alters protein conformation. Recently such effects have been taken care of by introducing a confining potential $-\frac{\partial}{\partial{R_n}}(\frac{k_c}{2}(R_n-0)^2)$ where, $k_c$ is the spring constant. In addition such a chain can posses what is known as the internal friction, $\xi_{int}$ and can be termed as ``Solvent Dependent Compacted Rouse with Internal Friction (SDCRIF)".  Earlier works discuss in detail how the inclusion of internal friction is done \cite{khatri2007, chakrabarti2013}. Like the Rouse model the end to end distribution of the polymer chain remains Gaussian in SDCRIF as well.  Experiments also showed that Gaussian Distributions of end to end distances of proteins are not too bad approximations \cite{moglich2, schuler2012}. Any deviations from the Gaussian behavior is negligibly small \cite{willmers1965}.  For SDCRIF the basic structure of the correlation function remains the same but has three extra parameters $\nu$, $\xi_{int}$ and $k_c$.

\begin{equation}
\left<X_{p\alpha}(0)X_{q\beta}(t)\right>=\frac{k_BT}{k_p^{SDCRIF}}\delta_{pq}\delta_{\alpha\beta}\exp\left(-t/\tau_p^{SDCRIF}\right)
\label{eq:xpcorrsd}
\end{equation}

Where, $k_p^{SDCRIF}=\frac{6 \pi^{2} k_B T p^{2\nu+1}}{N^{2\nu} b^2} + 2 N k_c$ and $\xi_p^{SDCRIF}=2 N \xi+ \frac{2 \pi^{2} p^{2\nu+1} \xi_{int}}{N^{2\nu}}$. The relaxation time for $p^{th}$ mode is $\tau_p^{SDCRIF}=\frac{\xi_p^{SDCRIF}}{k_p^{SDCRIF}}$.  As expected with $\nu=1/2$ which corresponds to $\theta$ solvent, SDCRIF gives back ``compacted Rouse with internal friction (CRIF)" \cite{chakrabarti2015}, when $k_c$ and $\xi_{int}$  both are nonzero.  A more realistic situation would be $\nu=3/5$ corresponding to good solvent. If the confining potential is removed, the model no longer represents a compacted chain but still remains solvent dependent Rouse and can be termed as ``solvent dependent Rouse with internal friction (SDRIF)". Further if the internal friction $\xi_{int}$ is ignored but $\nu\neq1/2$, the model reduces to solvent dependent Rouse (SDR), and as expected, in the limit when $\xi_{int}$, $k_c$ both are zero and $\nu=1/2$, Rouse chain is recovered. Another model would be ``solvent dependent compacted Rouse (SDCR)" when $\xi_{int}=0$, but $k_c\neq0$. SDCR can account for denaturant effect even in the absence of internal friction which we have used later to reproduce the reconfiguration time of an IDP in high denaturant concentration when internal friction should be negligible. Fig. (\ref{fig:a}) summarises the models and Table. \ref{tablerouse} depicts all the models along with the parameters describing the models.

\begin{table}[tbp]
\centering
\resizebox{\columnwidth}{!}{%
\begin{tabular}{|c||c|c|c|c|}
%\multicolumn{1}{r}{}
% &  \multicolumn{3}{c}{No Hydrodynamics} \\
\hline Parameter & SDR & SDRIF & SDCR & SDCRIF   \\ \hline
$\xi_p$ & $\xi_p^{SDR}=2N\xi$ & $\xi_p^{SDRIF}=\xi_p^{SDR}+\frac{2\pi^2p^{2\nu+1}\xi_{int}}{N^{2\nu}}$ & $\xi_p^{SDCR}=\xi_p^{SDR}$ & $\xi_p^{SDCRIF}=\xi_p^{SDRIF}$  \\ \hline
$k_p$ & $k_p^{SDR}=\frac{6 \pi^2 k_B T p^{2\nu+1}}{N^{2\nu} b^2}$ & $k_p^{SDRIF}=k_p^{SDR}$ & $k_p^{SDCR}=2Nk_c+k_p^{SDR}$& $k_p^{SDCRIF}=2Nk_c+k_p^{SDR}$ \\
\hline
$\tau_p$ & $\tau_p^{SDR}=\frac{\xi_p^{SDR}}{k_p^{SDR}}=\frac{\tau^{SDR}}{p^{2\nu+1}}$ & $\tau_p^{SDRIF}=\frac{\xi_p^{SDR}}{k_p^{SDRIF}}+\tau_{int}^{SDRIF}$ & $\tau_p^{SDCR}=\frac{\xi_p^{SDR}}{k_p^{SDCR}}$ & $\tau_p^{SDCRIF}=\frac{\xi_p^{SDR}}{k_p^{SDCRIF}}+\tau_{int}^{SDCRIF}$ \\
\hline
$\tau_{int}$ & $0$ & $\tau_{int}^{SDRIF}=\frac{\xi_{int}}{k}$ &0 & $\tau_{int}^{SDCRIF}=\frac{\xi_{int}}{k+k_c N^{2\nu+1}/p^{2\nu+1} \pi^2}$\\ \hline
\end{tabular}
}
%\vspace{0.1 in}
\caption{List of parameters for SDR, SDRIF, SDCR and SDCRIF}
\label{tablerouse}
\end{table}

\section{Calculation Methods}

\subsection{Reconfiguration time}

The time correlation function of the end-to-end vector is calculated from the correlation of normal modes as follows

\begin{equation}
{\phi_{N0}}(t)=\left<{R}_{N0}(0).{R}_{N0}(t)\right>^{SDCRIF}
=16 \sum\limits_{p=1}^\infty \frac{3 k_B T}{k_p^{SDCRIF}}exp(-t/\tau_p^{SDCRIF})
\label{eq:corr}
\end{equation}

Therefore,

\begin{equation}
{\phi_{N0}}(0)=\left<{R}_{N0}^2\right>^{SDCRIF}_{eq}
=16 \sum\limits_{p=1}^\infty \frac{3 k_B T}{k_p^{SDCRIF}}
\label{eq:corr1}
\end{equation}

\noindent The exact expression for $\left<R_{N0}^2\right>_{eq}^{SDCRIF}$ is not analytically trackable. However for $\theta$ solvent when $\nu=1/2$, it has an analytical expression \cite{chakrabarti2015}
\begin{equation}
\left<R_{N0}^2\right>_{eq,\nu=1/2}^{CRIF}=\frac{2 b \sqrt{3 k_B T}}{\sqrt{k_c}} tanh[\frac{N b \sqrt{k_c}}{2 \sqrt{3 k_B T}}]
\label{req}
\end{equation}

\noindent Fortunately for SDRIF,  $\left<R^2_{N0}\right>_{eq}$ has an analytical expression too

\begin{equation}
\left<R^2_{N0}\right>_{eq}^{SDRIF}=\frac{2^{2-2\nu}(-1+2^{1+2\nu})\zeta[1+2\nu]b^2 N^2  }{\pi^2}
\label{eq:rcorr}
\end{equation}

\noindent Where, $\zeta[x]=\frac{1}{\Gamma(x)}\int\limits_{0}^\infty\frac{u^{x-1}}{e^{u}-1}$ is Riemann Zeta function and $\Gamma(x)$ is the gamma function.

Reconfiguration time $\tau_{N0}$ is obtained by integrating the normalized $\phi_{N0}(t)$ \cite{chakrabarti2013, chakrabarti2014}

\begin{equation}
\tau_{N0}=\int\limits_{0}^\infty dt \tilde\phi_{N0}(t)
\label{eq:recon}
\end{equation}

\noindent Where, $\tilde\phi_{N0}(t)=\frac{{\phi_{N0}}(t)}{{\phi_{N0}}(0)}$

\subsection{Looping time}

Wilemski Fixman (WF) approach is a widely accepted method to calculate the time required to form a loop between two parts of a Gaussian polymer chain \cite{wilemski1974}. This formalism gives the following expression for the  looping time between two ends of the chain.

\begin{equation}
\tau_{N0}^{loop}=\int_0^{\infty} dt \left(\frac{C_{N0}(t)}{C_{N0}(\infty)}-1\right)
\label{eq:looping-time}
\end{equation}

\noindent Where, $C_{N0}(t)$ is the sink-sink correlation function given by

\begin{equation}
C_{N0}(t)=\int dR_{N0} \int dR_{N0,0} S(R_{N0})G(R_{N0},t | R_{N0,0},0)S(R_{N0,0})P(R_{N0,0})
\label{eq:sink-sink}
\end{equation}

\noindent $G(R_{N0},t | R_{N0,0},0)$ is the conditional probability of the polymer to have end-to-end distance $R_{N0}$ at time $t$, which was $R_{N0,0}$ at time $t=0$.

\begin{equation}
G(R_{N0},t | R_{N0,0},0)=\left(\frac{3}{2\pi \left<R^2_{N0}\right>_{eq}^{SDCRIF}}\right)^{3/2}\left(\frac{1}{(1-\tilde\phi_{N0}^2(t))^{3/2}}\right) exp\left[-\frac{3({R_{N0}}-\tilde\phi_{N0}(t)R_{N0,0})^2}{2\left<R^2_{N0}\right>_{eq}^{SDCRIF}(1-\tilde\phi_{N0}^2(t))}\right]
\label{eq:green}
\end{equation}

\noindent When the sink function $S(R_{mn})$ \cite{bagchibook, sebastianpra1992, debnath2006} is chosen to be a delta function then the looping time of SDCRIF has the following expression \cite{pastor}

\begin{equation}
\tau_{N0}^{loop,SDCRIF}=\int_0^{\infty} dt \left( \frac{exp[-2 \chi_0 \tilde\phi_{N0}^2(t)/(1-\tilde\phi_{N0}^2(t))]sinh[(2\chi_0\tilde\phi_{N0}(t))/(1-\tilde\phi_{N0}^2(t))]}{(2 \chi_0 \tilde\phi_{N0}(t))\sqrt{1-\tilde\phi_{N0}^2(t)}}-1\right)
\label{eq:taudeltamn}
\end{equation}

\noindent Where,

\begin{equation}
\chi_0=\frac{3 a^2}{2 \left<R_{N0}^2\right>_{eq}^{SDCRIF}}
\label{chi}
\end{equation}

\section{Results}

\subsection{Equilibrium end to end distribution}

\begin{figure}
\centering
 \includegraphics[width=0.8\textwidth]{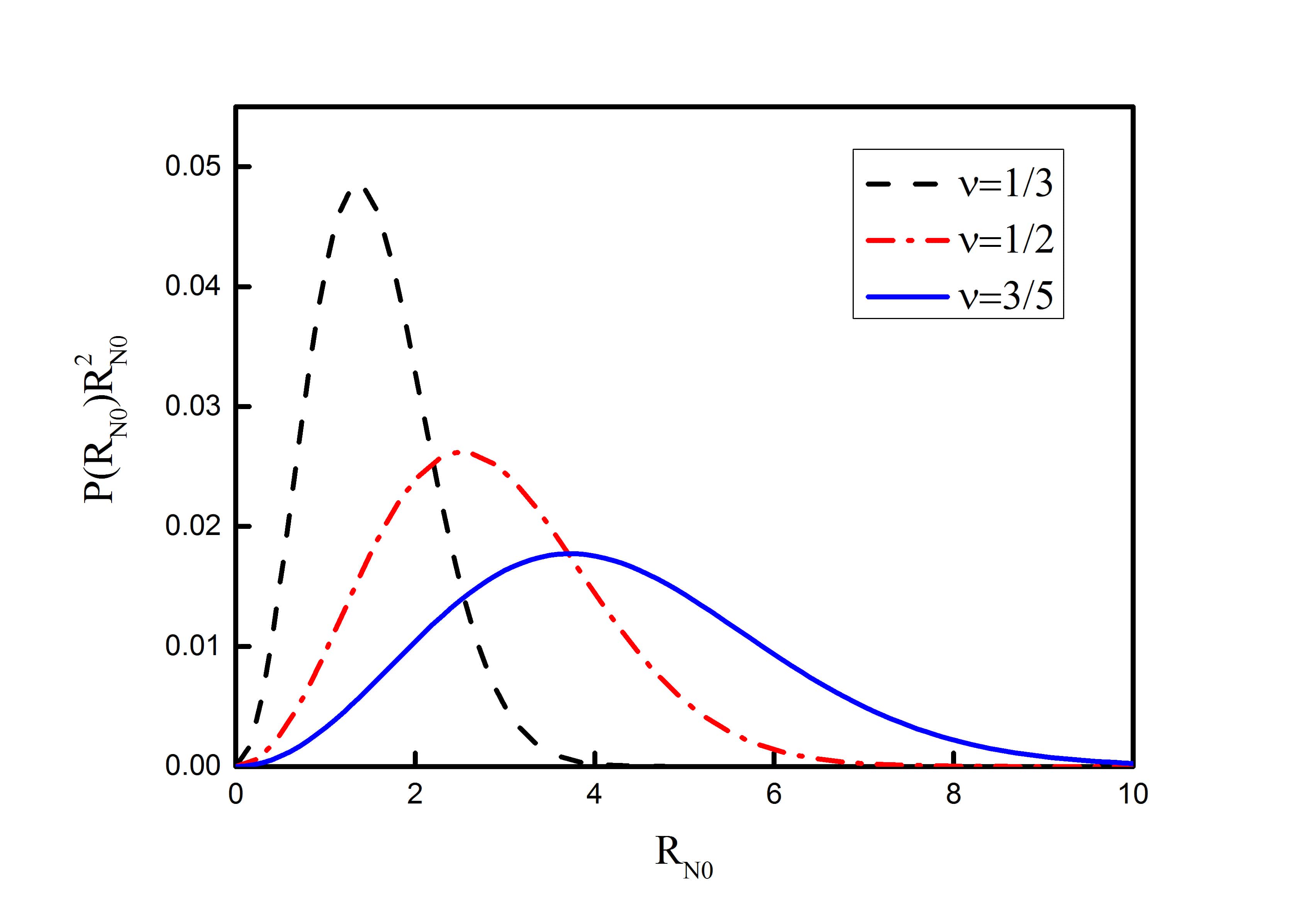}
 \caption{End-to-end distribution at different solvent quality with $k_c=0$. The values of parameters used are $N =66$, $b = 3.8\times10^{-10} m$, $\xi = 9.42\times10^{-12} kgs^{-1}$, $k_B=1.38\times10^{-23} JK^{-1}$ and $T=300K$.}
 \label{fig:b}
 \end{figure}

\begin{figure}
\centering
 \includegraphics[width=0.8\textwidth]{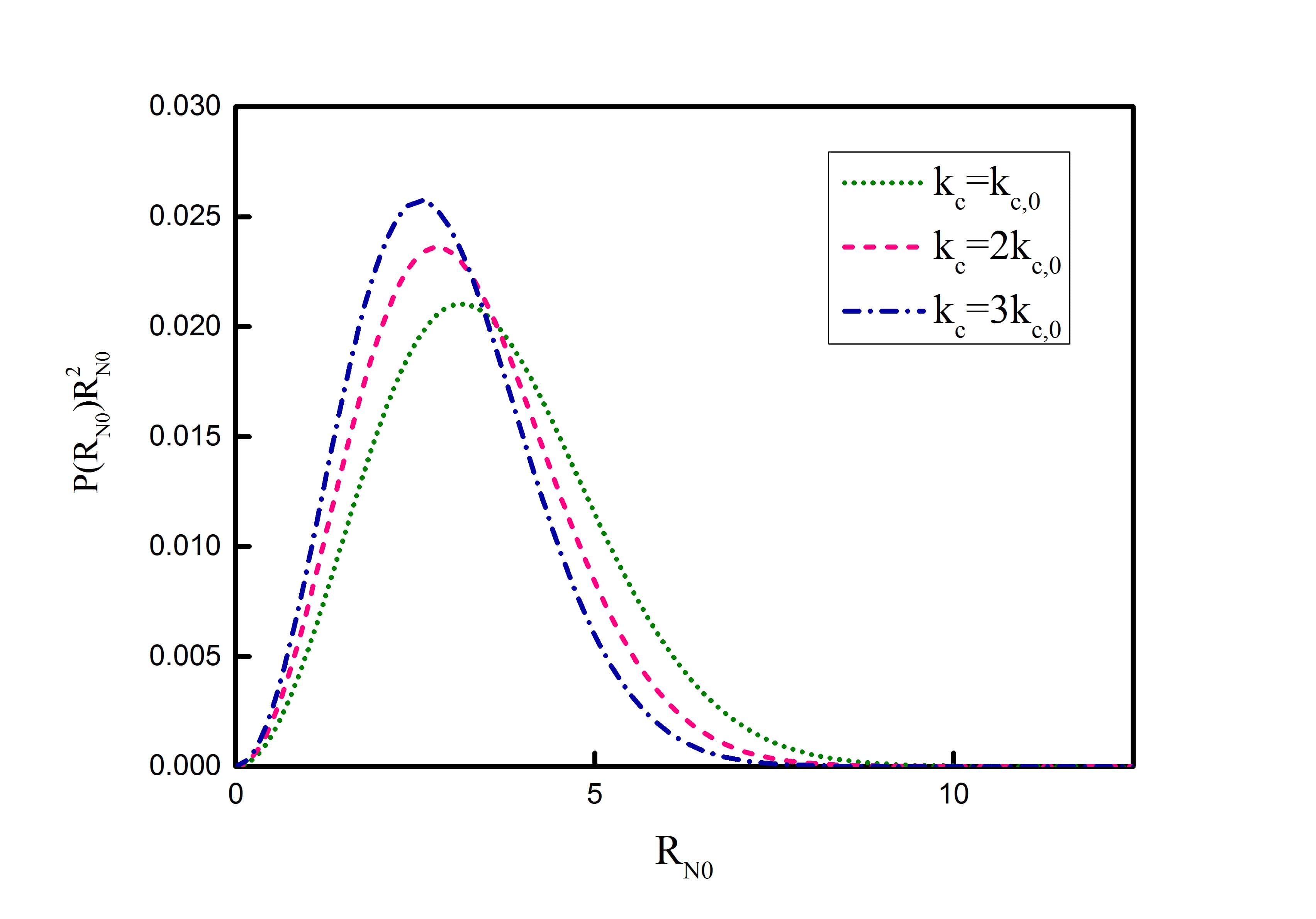}
 \caption{End-to-end distribution at different values of $k_c$ for good solvent ($\nu=3/5$). The values of parameters used are $N =66$, $\tilde{k}_c=0$, $b = 3.8\times10^{-10} m$, $\xi = 9.42\times10^{-12} kgs^{-1}$, $k_B=1.38\times10^{-23} JK^{-1}$ and $T=300K$.}
 \label{fig:c}
 \end{figure}

The equilibrium distribution of the vector connecting end-to-end monomers of a Gaussian chain is given by $P(R_{N0})=\left(\frac{3}{2\pi\left<R^2_{N0}\right>}\right)^{3/2}exp\left[-\frac{3R^2_{N0}}{2\left<R^2_{N0}\right>}\right]$, where, $\left<R^2_{N0}\right>$ denotes the average equilibrium end to end distance of the polymer. This expression holds for SDCRIF as well since the model is considered to be Gaussian in our description. In Fig. (\ref{fig:b}) the equilibrium distribution of the end-to-end distance is shown for the polymer at different solvent quality in absence of the confining potential. As expected the distribution plot is broader for the polymer in good solvent $(\nu=3/5)$ in comparison to the polymer in poor solvent ($\nu=1/3$). From good to $\theta$ to bad solvent the most probable end to end distance or the peak position shits to lower value as a signature of swelled to collapsed transition. The parameters have been chosen in accordance to the work by Schuler's group \cite{schuler2012} on cold shock protein (Csp) and prothymosin $\alpha$ (ProT$\alpha$) and has been mentioned in all the figures. In Fig. (\ref{fig:c}) the same distribution is shown for good solvent but in presence of three different values of $k_c$. As the $k_c$ controls the compactness of the polymer, increasing the value of $k_c$ should result in the higher degree of internal friction. Since there is no first principle relation between $k_c$ and $\xi_{int}$, we have used an ansatz in our calculation \cite{chakrabarti2015, bhuyan2013}. The ansatz connects $k_c$ with $\xi_{int}$ as follows $k_c=\tilde{k}_c+k_{c,0} (c_0+c_1n_b+c_2n_b^2+....)$, where, $k_{c,0}=A\frac{\xi_{int,0}}{\tau_{int,0}}$, and $\xi_{int}=\xi_{int,0}(c_0+c_1n_b+c_2n_b^2+....)$. $\xi_{int,0}$ is the zeroth order approximation to internal friction which only accounts for the interactions between a monomer with its two nearest neighbours, $\tau_{int,0}$ is the corresponding time scale, $n_b$ is the number of non-adjacent monomers contributing to the internal friction and $A$ is considered typically in the order of $\pi^2/N^{2\nu+1}$ whereas other parameters are constants with no or very weak dependence on chain properties. Therefore $k_c$ has a part proportional to $\xi_{int, 0}$. However, in the limit $\xi_{int, 0}=0$, $k_c=\tilde{k}_c$. If the value of $k_c$ or the strength of the confinement increases the end to end distance probability distribution becomes narrower, a sign of more compacted polymer chain.  This can clearly be seen from the plot of end to end distance distribution at different values of $k_c$ depicted in Fig. (\ref{fig:c}). Thus $k_c$ and $\nu$ plays significant roles in deciding the width and height of the end to end distribution. But the the parameters have different physical significance and play different roles in the chain dynamics. The exponent $\nu$ takes care of the quality of the solvent around the chain and has no dependence on $\xi_{int}$ and no or very weak dependence on denaturant concentration \cite{eatonbiophysj2006}. But $k_c$ on the other hand takes care the effect of the denaturant on the chain and has $\xi_{int}$ dependence as well. Here we also attempt to make a qualitative comparison of our model SDCRIF with the very recent simulation from Luo's \cite{luo2015} group. Fig. 3. of \cite{luo2015}, shows similar trend of the end to end probability distribution on changing the parameter $\lambda$, a measure of attractive interaction between the beads. Higher the value of $\lambda$ poorer the solvent, narrower the distribution.

\subsection{Relaxation time for the Normal modes}

\begin{figure}
\centering
 \includegraphics[width=0.8\textwidth]{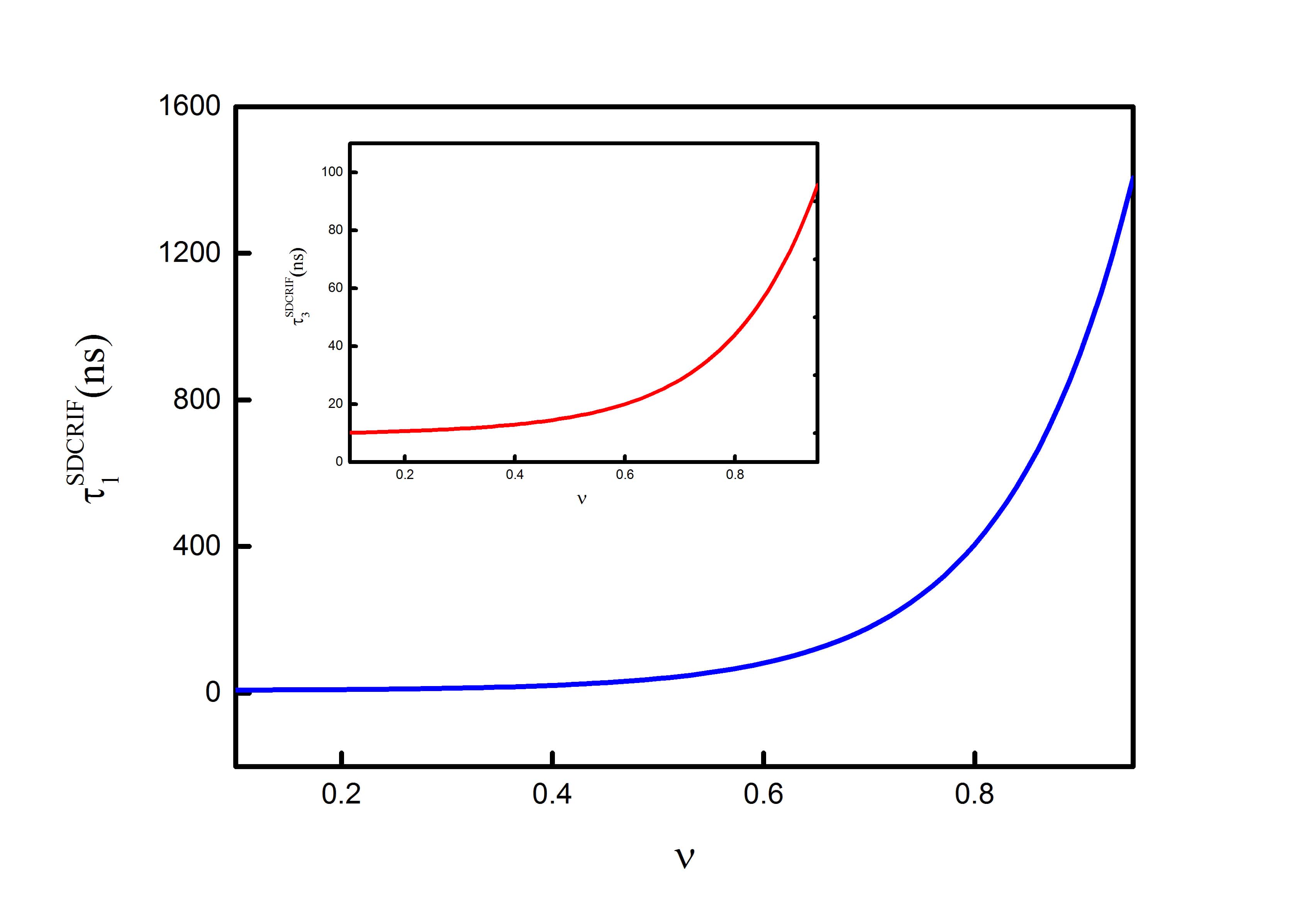}
 \caption{$\tau_p^{SDCRIF}$ vs $\nu$ for $p=1$ and $p=3$ (inset). The values of parameters used are $N =66$, $k_c=k_{c,0}$, $\tilde{k}_c=0$, $b = 3.8\times10^{-10} m$, $\xi = 9.42\times10^{-12} kgs^{-1}$, $\xi_{int,0} = 100\times\xi$, $k_B=1.38\times10^{-23} JK^{-1}$ and $T=300K$.}
 \label{fig:d}
 \end{figure}

\begin{figure}
\centering
 \includegraphics[width=0.8\textwidth]{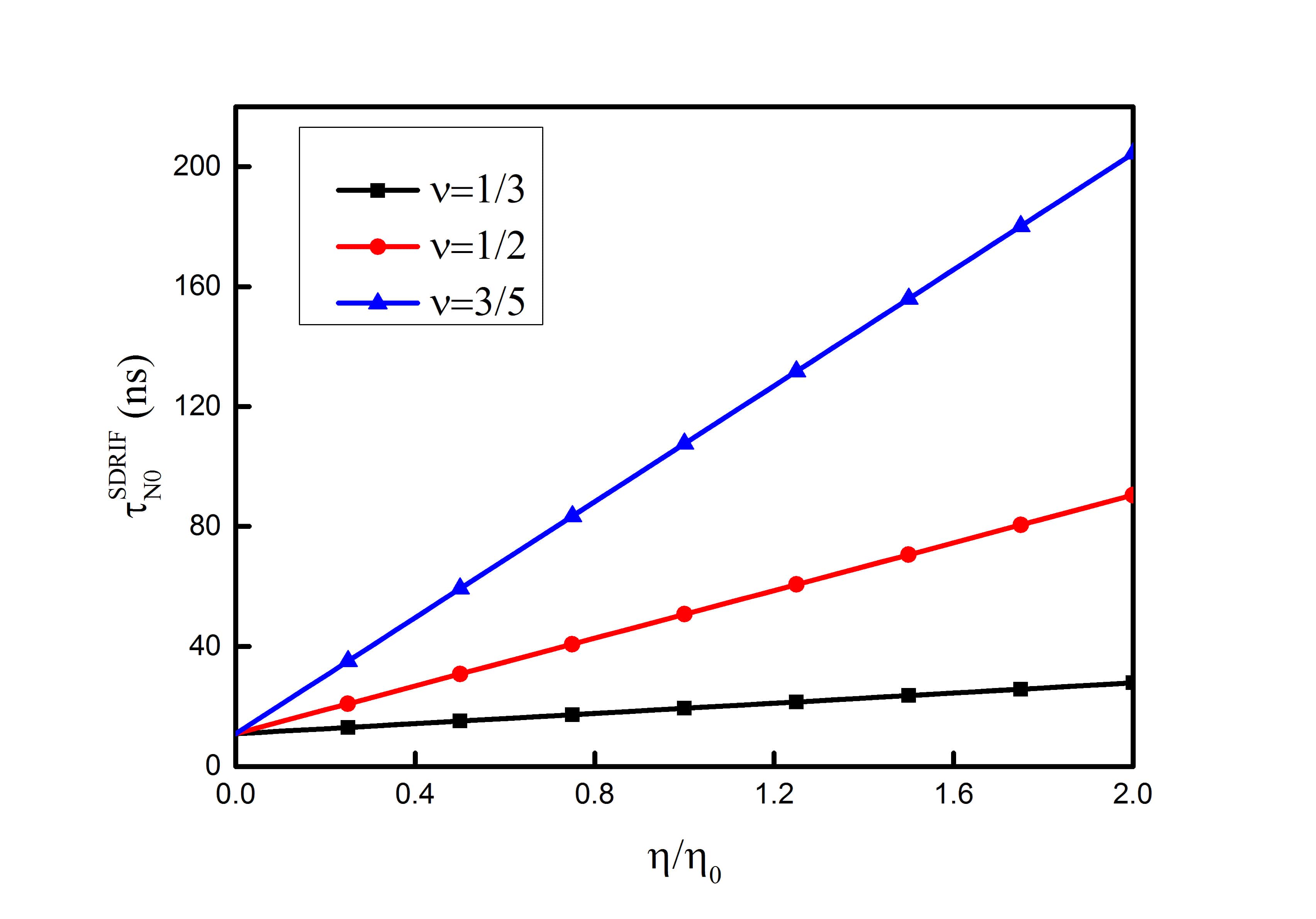}
 \caption{Reconfiguration time $(\tau_{N0}^{SDRIF})$ vs solvent viscosity at different solvent quality. The values of parameters used are $N =66$, $k_c=0$, $b = 3.8\times10^{-10} m$, $\xi = 9.42\times10^{-12} kgs^{-1}$, $\xi_{int,0} = 100\times\xi$, $k_B=1.38\times10^{-23} JK^{-1}$ and $T=300K$.}
 \label{fig:e}
 \end{figure}

In order to calculate reconfiguration time one has to calculate the correlation function between normal modes defined in Eq. \ref {eq:xpcorrsd}.  The functional from of $\tau_{p}^{SDCRIF}$ is given in Table. \ref{tablerouse}.  The lower normal modes contribute largely to looping and reconfiguration dynamics \cite{doi}. A plot of $\tau_{p}^{SDCRIF}$ vs $\nu$ for the first normal mode is shown in Fig. (\ref{fig:d}). This shows how $\tau_{p}^{SDCRIF}$ changes with the solvent quality if all the other parameters remain fixed. The plot shows that $\tau_{p}^{SDCRIF}$ (for $p=1$) increases with $\nu$. Thus for poor solvents corresponding to lower value of $\nu$, $\tau_{p}^{SDCRIF}$ is lower and as one approaches good solvent $\tau_{p}^{SDCRIF}$ increases resulting in  slower relaxation. This would definitely lead to slower reconfiguration and looping time in good solvent as compared to bad or poor solvent.  Similar trend for the third normal mode can be seen from the plot in the inset of the Fig. (\ref{fig:d}). Recent simulation also confirm this \cite{luo2015} and shows an order of magnitude increase in the looping time in the good solvent as compared to the poor solvent. The reconfiguration times of a chain for different values of $\nu$ are shown in Fig. (\ref{fig:e}). It can be seen that at a value $\eta/\eta_0=1$, the reconfiguration time increases by a factor of $6$ on changing $\nu=1/3$ to $\nu=3/5$.

 \begin{figure}
\centering
 \includegraphics[width=0.8\textwidth]{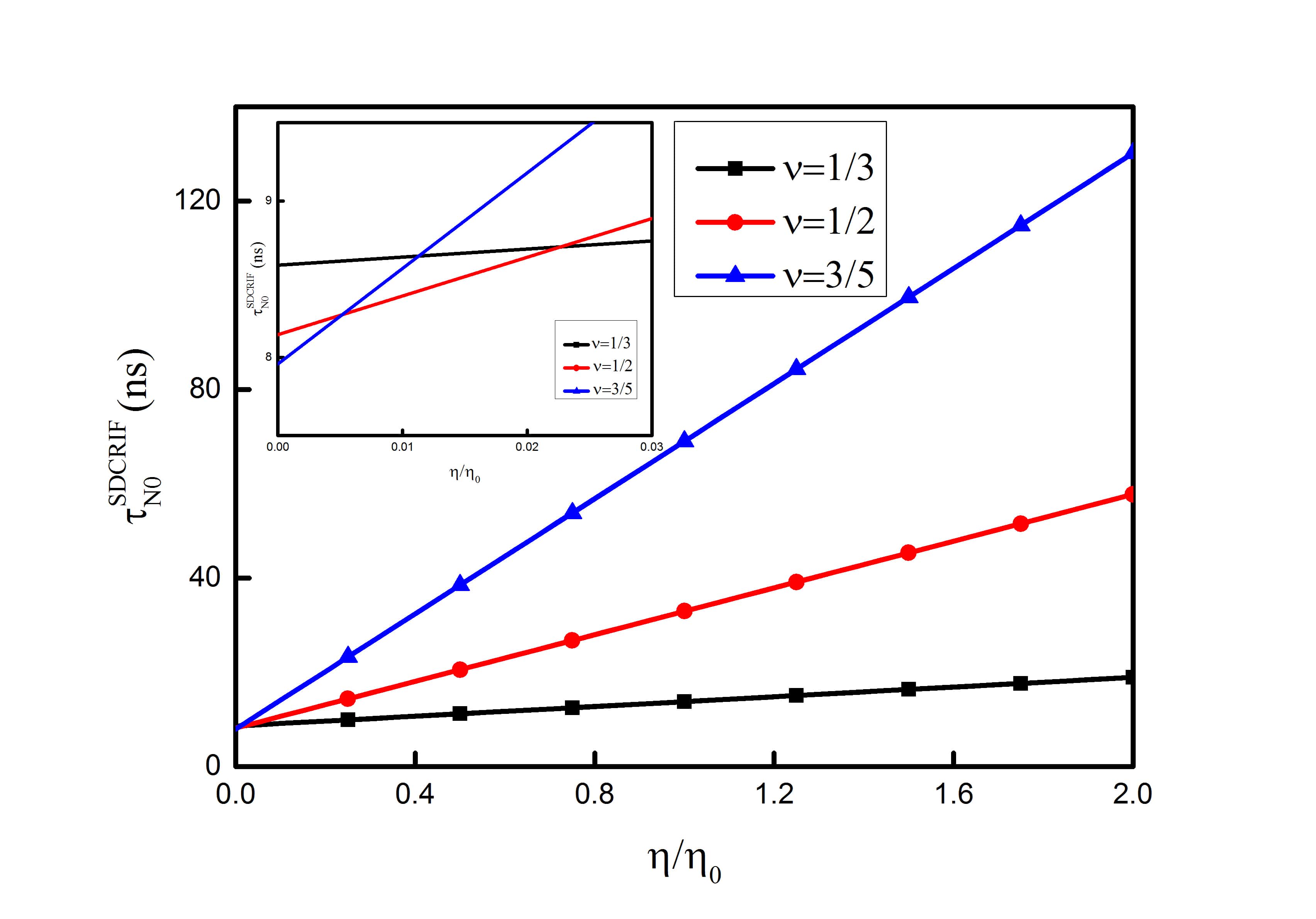}
 \caption{Reconfiguration time $(\tau_{N0}^{SDCRIF})$  vs solvent viscosity at different solvent quality. The values of parameters used are  $N =66$, $k_c=k_{c,0}$, $\tilde{k}_c=0$, $b = 3.8\times10^{-10} m$, $\xi = 9.42\times10^{-12} kgs^{-1}$, $\xi_{int,0} = 100\times\xi$, $k_B=1.38\times10^{-23} JK^{-1}$ and $T=300K$.}
 \label{fig:f}
 \end{figure}

  \begin{figure}
\centering
 \includegraphics[width=0.8\textwidth]{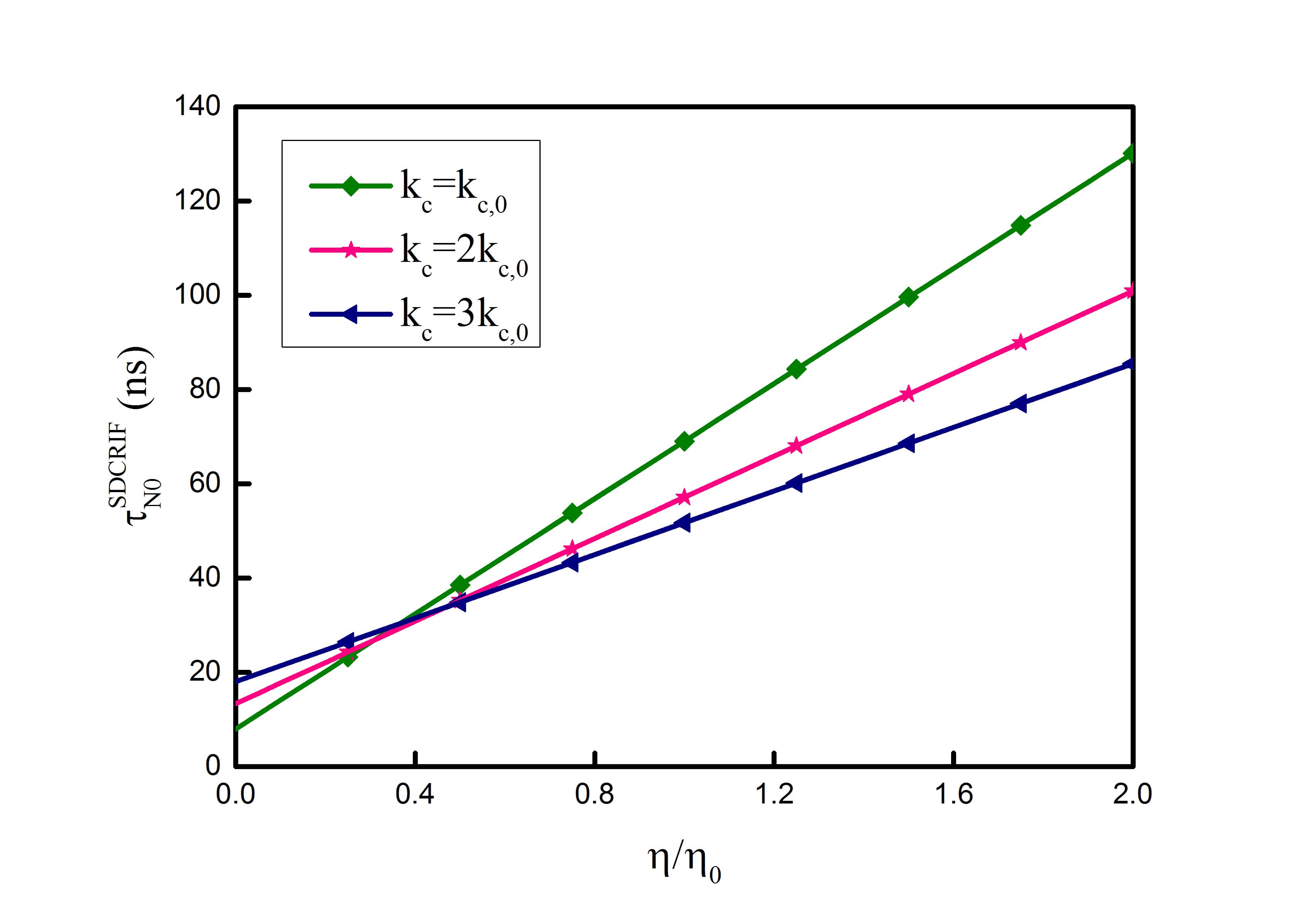}
 \caption{Reconfiguration time $(\tau_{N0}^{SDCRIF})$ vs solvent viscosity at different values of $k_c$ for good solvent ($\nu=3/5$). The values of parameters used are $N =66$, $\tilde{k}_c=0$, $b = 3.8\times10^{-10} m$,
 $\xi =9.42\times10^{-12} kgs^{-1}$, $k_B=1.38\times10^{-23} JK^{-1}$ and $T=300K$.}
 \label{fig:g}
 \end{figure}

\subsection{Solvent viscosity dependence}

One aspect of the reconfiguration time is its dependence on solvent viscosity \cite{chakrabarti2013}. Experiments show a linear dependence of reconfiguration time on solvent viscosity \cite{schuler2012}. In Fig. (\ref{fig:e}) the reconfiguration times  $(\tau_{N0}^{SDRIF})$ are plotted against the solvent viscosity at different values of $\nu$ in absence of any confining potential ($k_c=0$) and non zero internal friction ($\xi_{int} \neq 0$). The figure clearly shows $\tau_{N0}^{SDRIF}$ increases  linearly with increasing normalized solvent viscosity $\eta/\eta_0$, with a positive intercept in the limit $\eta/\eta_0\rightarrow0$, $\eta_0$ being the viscosity of water. What is surprising is the value of intercept is practically independent of solvent quality. This intercept corresponds to the internal friction which is completely dry in this case.  The trend can be fitted with an almost analytically exact expression $\tau_{N0}^{SDR}\simeq \tau_{int}^{SDR}+ C(\nu) \tau^{SDR}$, where $\tau^{SDR}=\frac{\xi N^{2\nu+1}b^2}{3\pi^2k_BT}$ and $C(\nu)$s have $\nu$ dependence and are $0.73$, $0.82$ and $0.87$ for the poor, $\theta$ and good solvent respectively. Thus poorer the solvent lower the slope. In poor solvent reconfiguration time is not only fast but also less affected by the viscosity of the solvent. This is presumably because of the collapsed form of the polymer chain. Fig. (\ref{fig:f}) is the plot of $\tau_{N0}^{SDCRIF}$ vs $\eta/\eta_0$ in presence of a fixed confining potential $k_{c,0}$ but at three different values of $\nu$ covering a range of solvent qualities. Interestingly in this case the intercepts are found to be different when the intercepts were magnified as can be seen in the inset of Fig. (\ref{fig:f}). This is because $\tau_{int}^{SDCRIF}$ has $k_c$ as well as $\nu$ dependence as can be seen from the Table. \ref{tablerouse}.  So even with the same value of  $k_c$, changing $\nu$ would result a change in the effective force constant, $k_{eff}\simeq k+k_c N^{2\nu+1}/\pi^2p^{2\nu+1}$ of the chain and hence a change in the value of the intercept. In all the cases $\xi_{int}$ remains the same and the time scales associated with the internal friction $\tau_{int}^{SDCRIF}$  are changed by very small amount which means the dependence of $\tau_{int}^{SDCRIF}$ on $\nu$ is very weak and it cannot reproduce the denaturant effect observed in the experiments, which is why we need $k_c$ and the ansatz which connects $k_c$ with $\xi_{int}$ to take care of the denaturant effect on a polymer without changing the solvent quality. This can be seen in Fig.  (\ref{fig:g}) where we have looked into the viscosity dependence of $\tau_{N0}^{SDCRIF}$ in different values of $k_c$  for good solvent ($\nu=3/5$). In this case the change in the values of the intercepts is very evident because of the different degrees of compactness of the chain. Another important observation is the case with higher $k_c$ value has a higher intercept but it is less steeper making the change in the reconfiguration time less prone to the viscosity of the solvent.  This is how $k_c$ can replicate the experimental results \cite{schuler2012} where it has been found that if denaturant is introduced to a polymer solution, the polymer becomes less compact which results in lower value of internal friction $\xi_{int}$. $k_c$ controls the compactness of the protein thus connects to the denaturant concentration. The higher the denaturant concentration lower the $k_c$ value.  This is also reflected in our ansatz as increasing denaturant concentration would mean smaller $n_b$ and lower internal friction $\xi_{int}$. But $k_c$ does not speak for the solvent quality rather it is $\nu$ which accounts for that.  In the later section we used SDCRIF to compare experimental data on cold shock protein and prothymosin $\alpha$ (ProT$\alpha$) \cite{schuler2012} and have confirmed this. We have also looked at the viscosity dependence of the looping time and have found it to be $\sim \eta^\beta$ with $\beta<2$ \cite{bagchi2001, chakrabarti2013}. A detailed study on the viscosity dependence of the looping time can be found in one of our earlier works \cite{chakrabarti2013}.

\subsection{Chain length dependence}

\begin{figure}
\centering
 \includegraphics[width=0.8\textwidth]{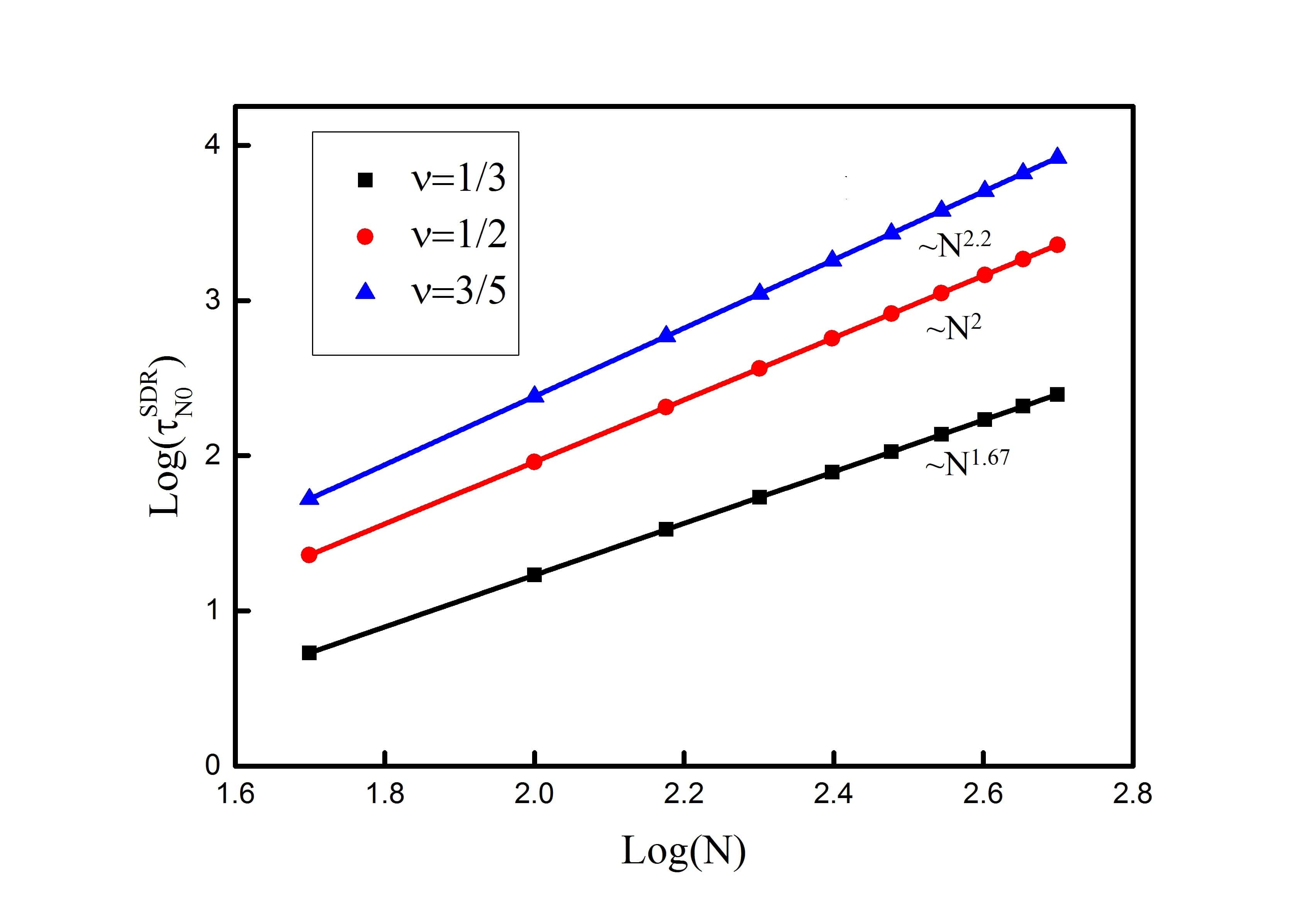}
 \caption{$Log(\tau_{N0}^{SDR})$ vs $Log(N)$ at different solvent quality. The values of parameters used are $k_c=0$, $b = 3.8\times10^{-10} m$, $\xi = 9.42\times10^{-12} kgs^{-1}$, $k_B=1.38\times10^{-23} JK^{-1}$, $\xi_{int}=0$ and $T=300K$.}
 \label{fig:h}
 \end{figure}

\begin{figure}
\centering
 \includegraphics[width=0.8\textwidth]{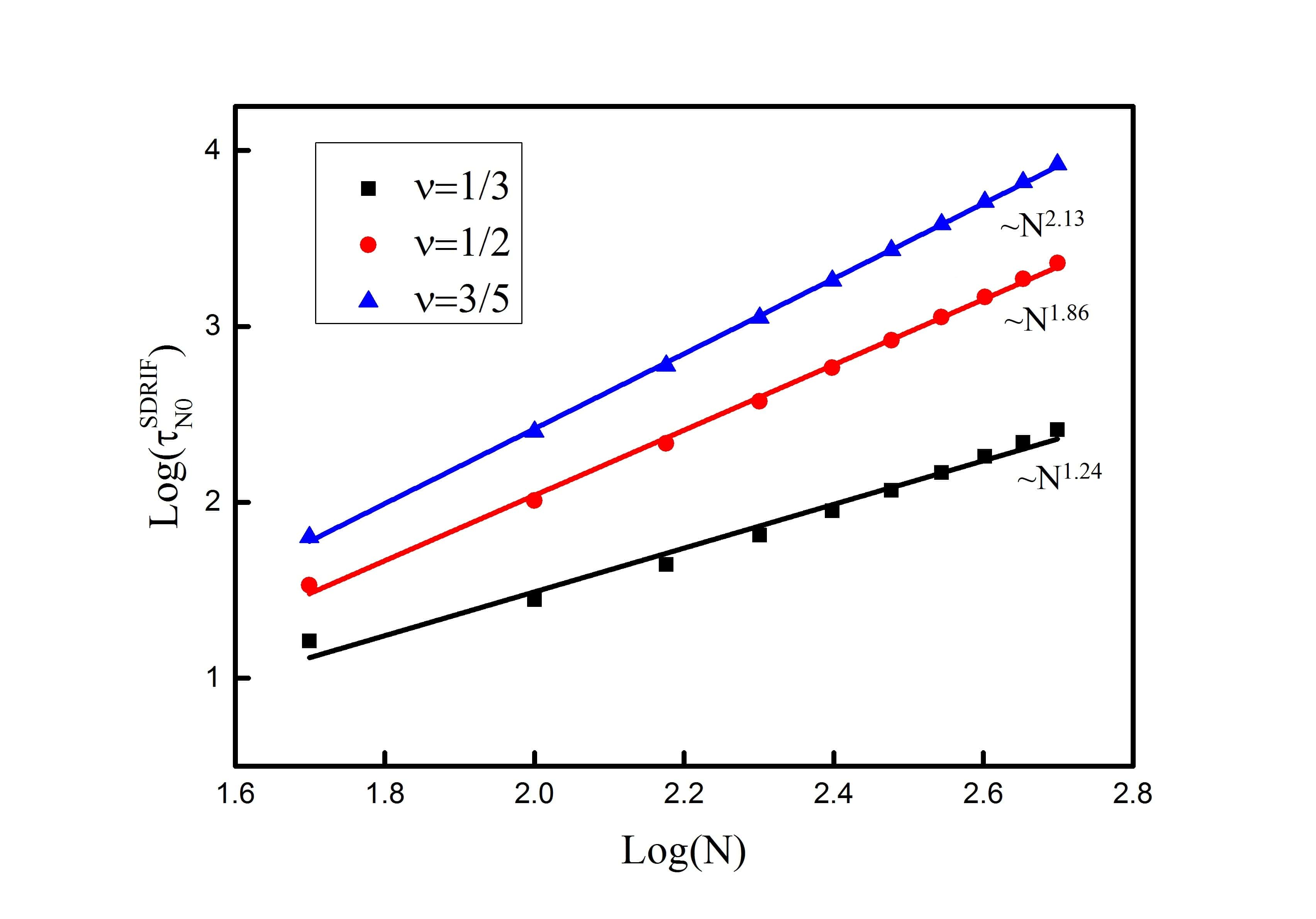}
 \caption{$Log(\tau_{N0}^{SDRIF})$ vs $Log(N)$ at different solvent quality. The values of parameters used are  $k_c=0$, $b = 3.8\times10^{-10} m$, $\xi = 9.42\times10^{-12} kgs^{-1}$, $\xi_{int,0} = 100\times\xi$, $k_B=1.38\times10^{-23} JK^{-1}$ and $T=300K$.}
 \label{fig:i}
 \end{figure}

\begin{figure}
\centering
 \includegraphics[width=0.8\textwidth]{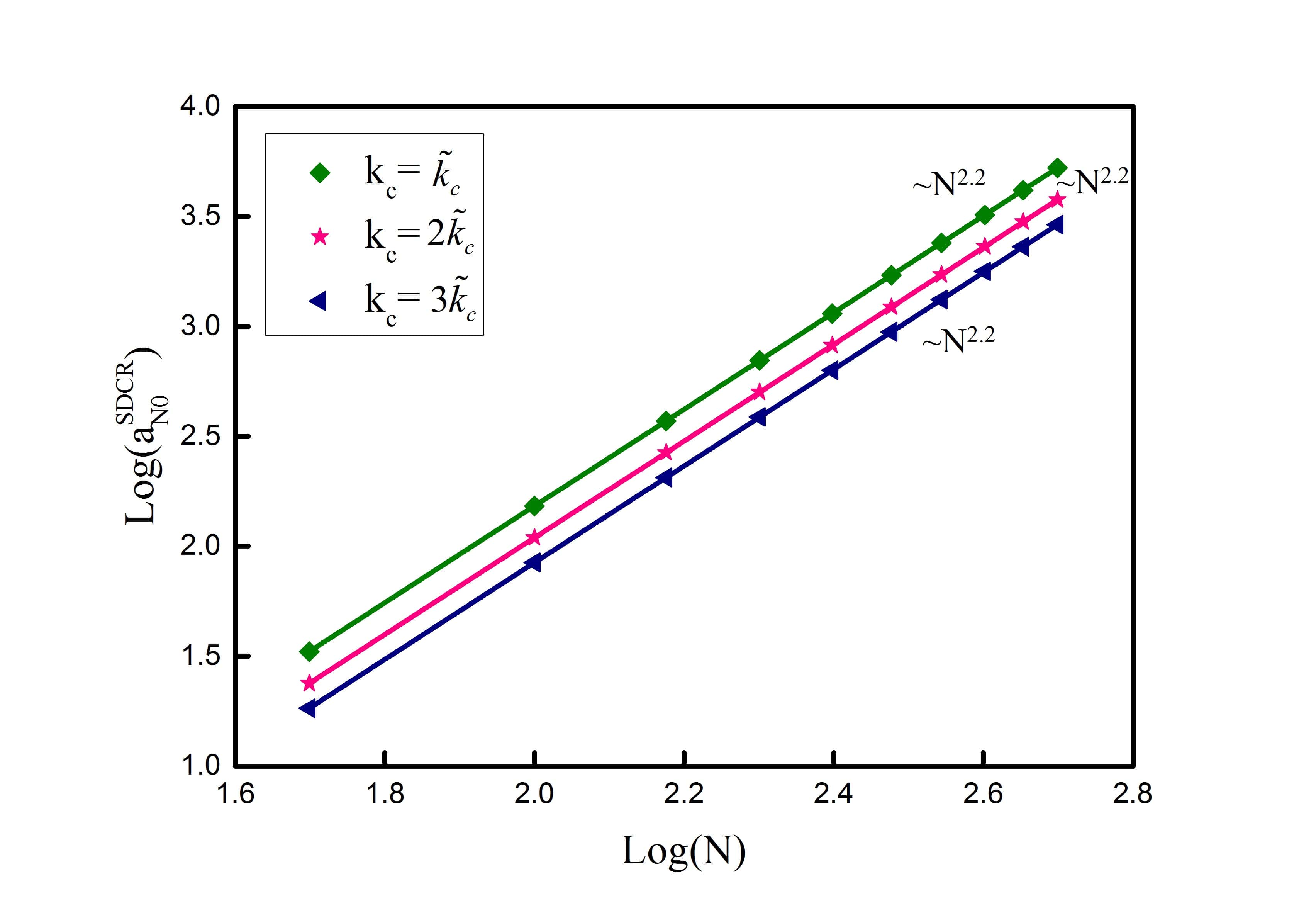}
 \caption{$Log(\tau_{N0}^{SDRCRIF})$ vs $Log(N)$ at different values of $k_c$ for good solvent ($\nu=3/5$). The values of parameters used are $b = 3.8\times10^{-10} m$, $\xi = 9.42\times10^{-12} kgs^{-1}$, $\xi_{int,0} = 0$, $k_B=1.38\times10^{-23} JK^{-1}$ and $T=300K$.}
 \label{fig:j}
 \end{figure}

\begin{figure}
\centering
 \includegraphics[width=0.8\textwidth]{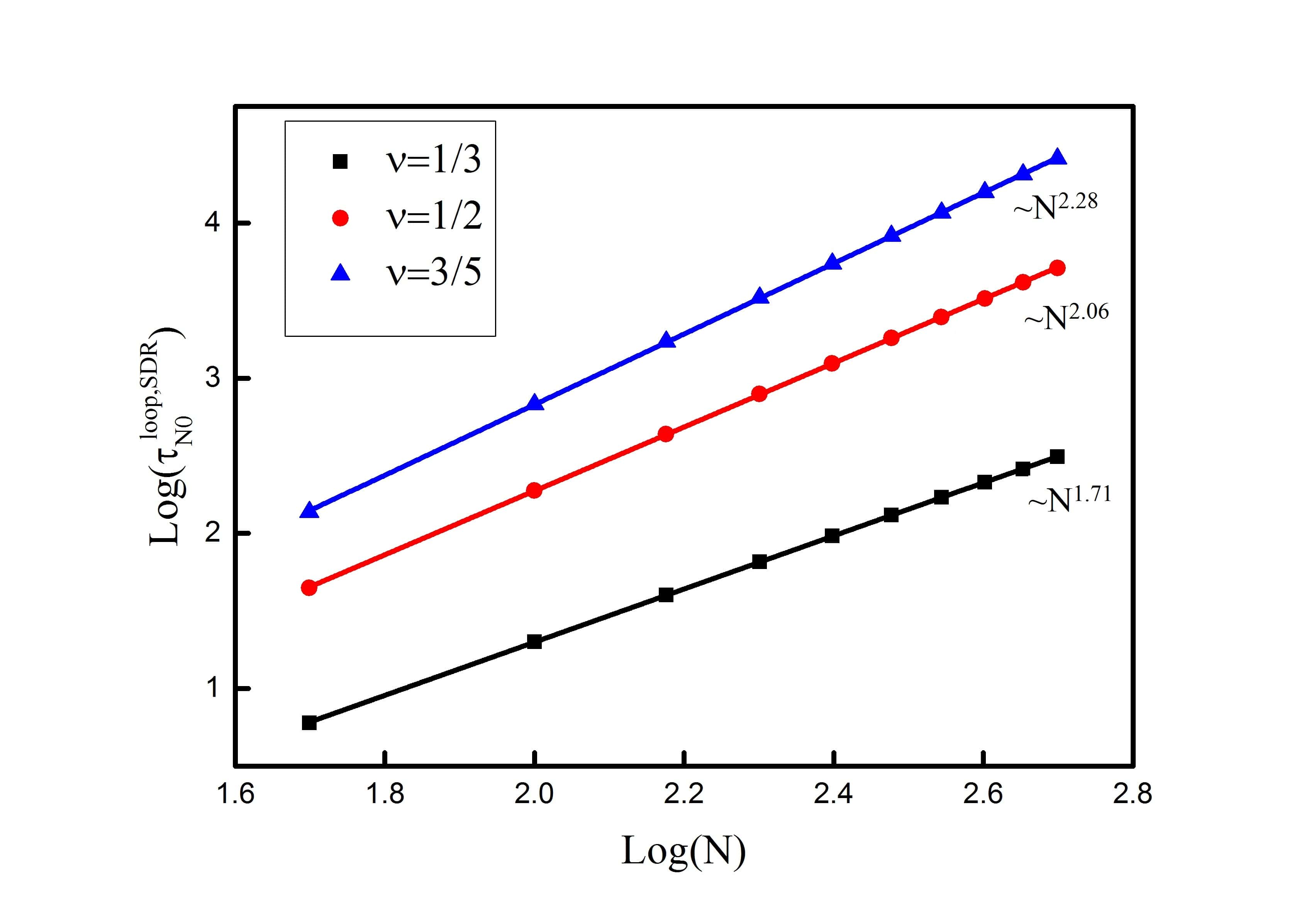}
 \caption{$Log(\tau_{N0}^{loop,SDR})$ vs $Log(N)$ at different solvent quality. The values of parameters used are $k_c=0$, $b = 3.8\times10^{-10} m$, $\xi = 9.42\times10^{-12} kgs^{-1}$, $\xi_{int,0} = 0$, $k_B=1.38\times10^{-23} JK^{-1}$ and $T=300K$.}
 \label{fig:k}
 \end{figure}

\begin{figure}

\centering
 \includegraphics[width=0.8\textwidth]{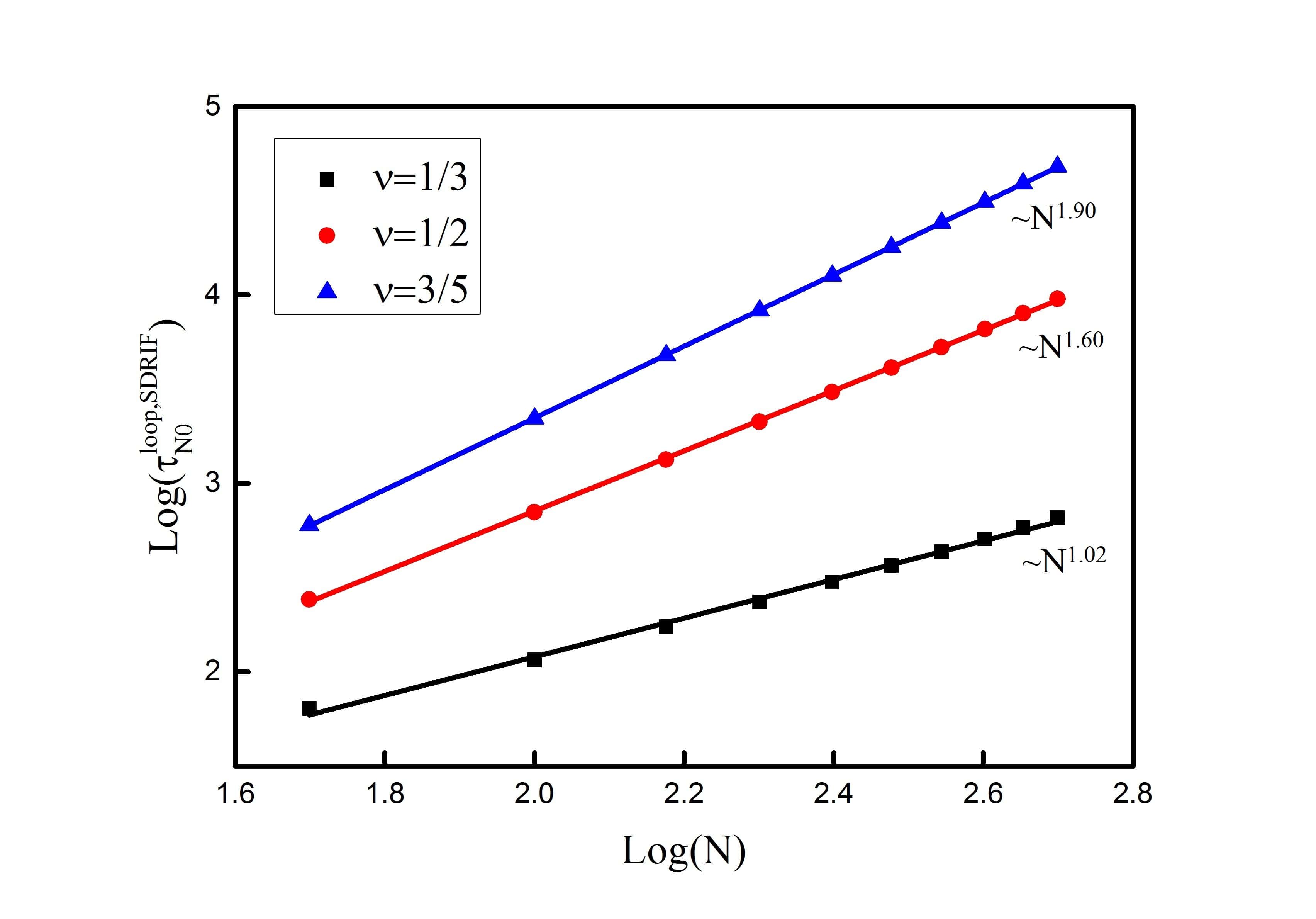}
 \caption{$Log(\tau_{N0}^{loop,SDR})$ vs $Log(N)$ at different solvent quality. The values of parameters used are $k_c=0$, $b = 3.8\times10^{-10} m$, $\xi = 9.42\times10^{-12} kgs^{-1}$, $\xi_{int,0} = 100\times\xi$, $k_B=1.38\times10^{-23} JK^{-1}$ and $T=300K$.}
 \label{fig:l}
 \end{figure}

Fig. (\ref{fig:h}) is the log-log plot of the reconfiguration time for a chain with $k_c=0, \xi_{int}=0$ vs chain length $N$ of the polymer for different values of $\nu$, the parameter accounting for the solvent quality. There is a general trend, $\tau_{N0}^{SDR}\sim N^{2{\nu}+1}$ as $\tau_{N0}^{SDR}=C(\nu)\tau^{SDR}$.  This result is expected and can be predicted by looking at the scaling of the end to end vector correlation function with the chain length. When the same calculations were done in presence of internal friction as shown in Fig. (\ref{fig:i}) the dependence on the chain length becomes weaker in general but the relative trend remains the same, in poor solvent dynamics is faster. Although the chain reconfiguration in general become slower due to the presence of internal friction. To see how the chain dynamics changes in presence of the confining potential the same calculations were again performed for three different values of $k_c$ but putting internal friction $\xi_{int}=0$ in good solvent ($\nu=3/5$) as shown in Fig. (\ref{fig:j}). Here as $\xi_{int}=0$ therefore, $k_{c,0}=0$ and $k_c=\tilde{k}_c$ whereas $\tilde{k}_c$ is scaled as $k_{c,0}$. This result is very interesting as it can be seen from the Fig. (\ref{fig:j}), that the $N$ dependence remains practically unchanged. The reason for this as follows. We restrict ourselves to a value of $k_c$  such that the effective force constant $k_{eff}\simeq k+k_c N^{2\nu+1}/\pi^2p^{2\nu+1}$ is in the same order of magnitude of $k$. For example a choice of $k_c\simeq k \pi^2 /N^{2\nu+1}$  gives $k_{eff}\simeq k(1+p^{2\nu+1})$. Now since lower normal modes contribute mostly in reconfiguration dynamics \cite{doi}, taking the contribution from $p=1$ normal mode gives $k_{eff}\simeq 2k$ and thus leads to unchanged scaling of the reconfiguration time with the chain length $N$ for a given value of $\nu$. Similar calculations have been done for looping time  based on the WF approximation \cite{wilemski1974} as can be seen from Fig. (\ref{fig:k}) and Fig. (\ref{fig:l}). The same arguments hold for the looping time and the trends are similar. Scaling of the looping time as $N^2$ for the Rouse chain \cite{friedman, chakrabartiphysica1} as obtained earlier can be confirmed with a choice of  $\nu=1/2$, $k_c=0$ and $\xi_{int}=0$ and is shown in Fig. (\ref{fig:k}).


\subsection{Comparison  with experiments}

\begin{figure}
\centering
  \begin{tabular}{@{}cccc@{}}
    \includegraphics[width=.4\textwidth]{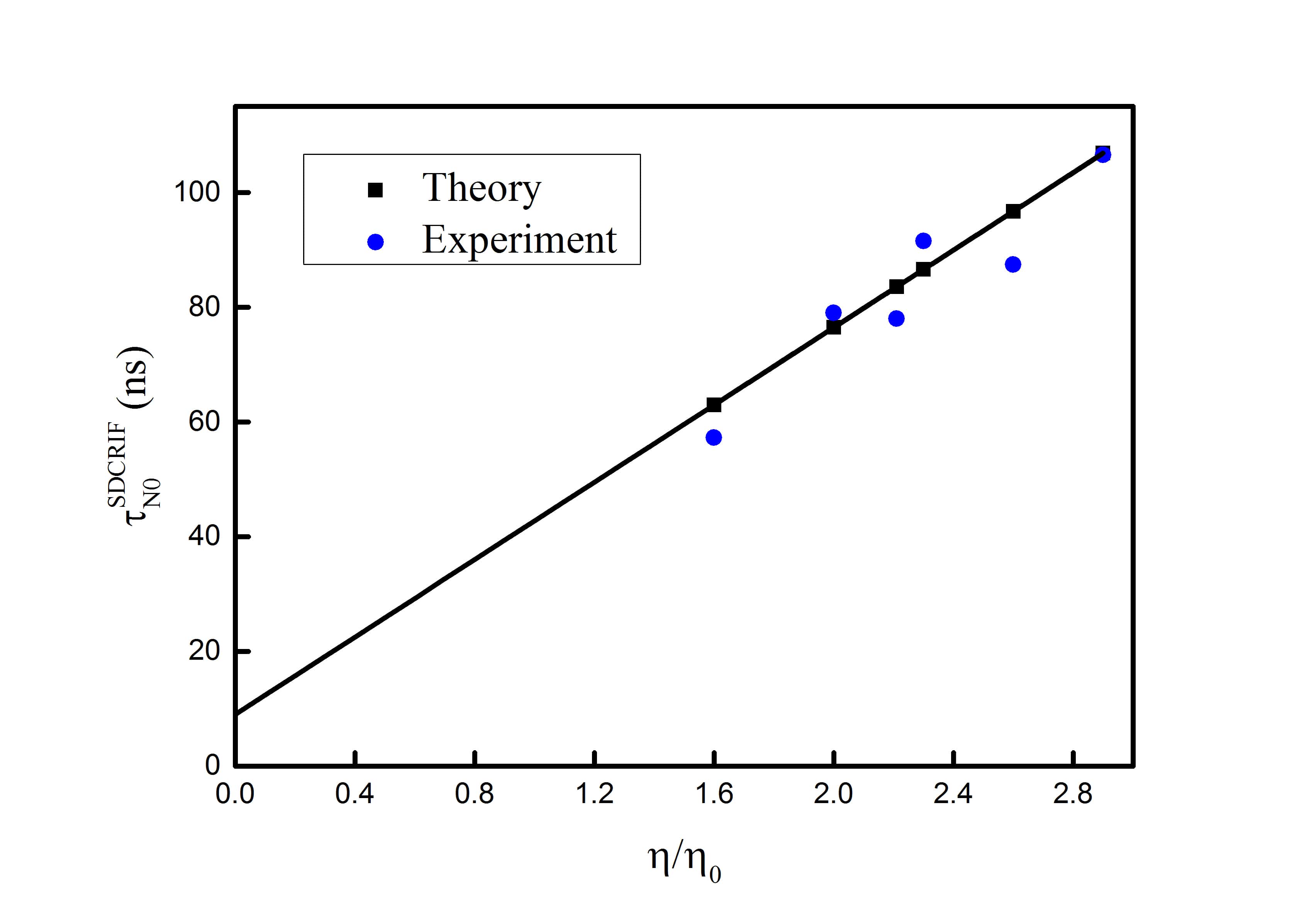}(a) &
    \includegraphics[width=.4\textwidth]{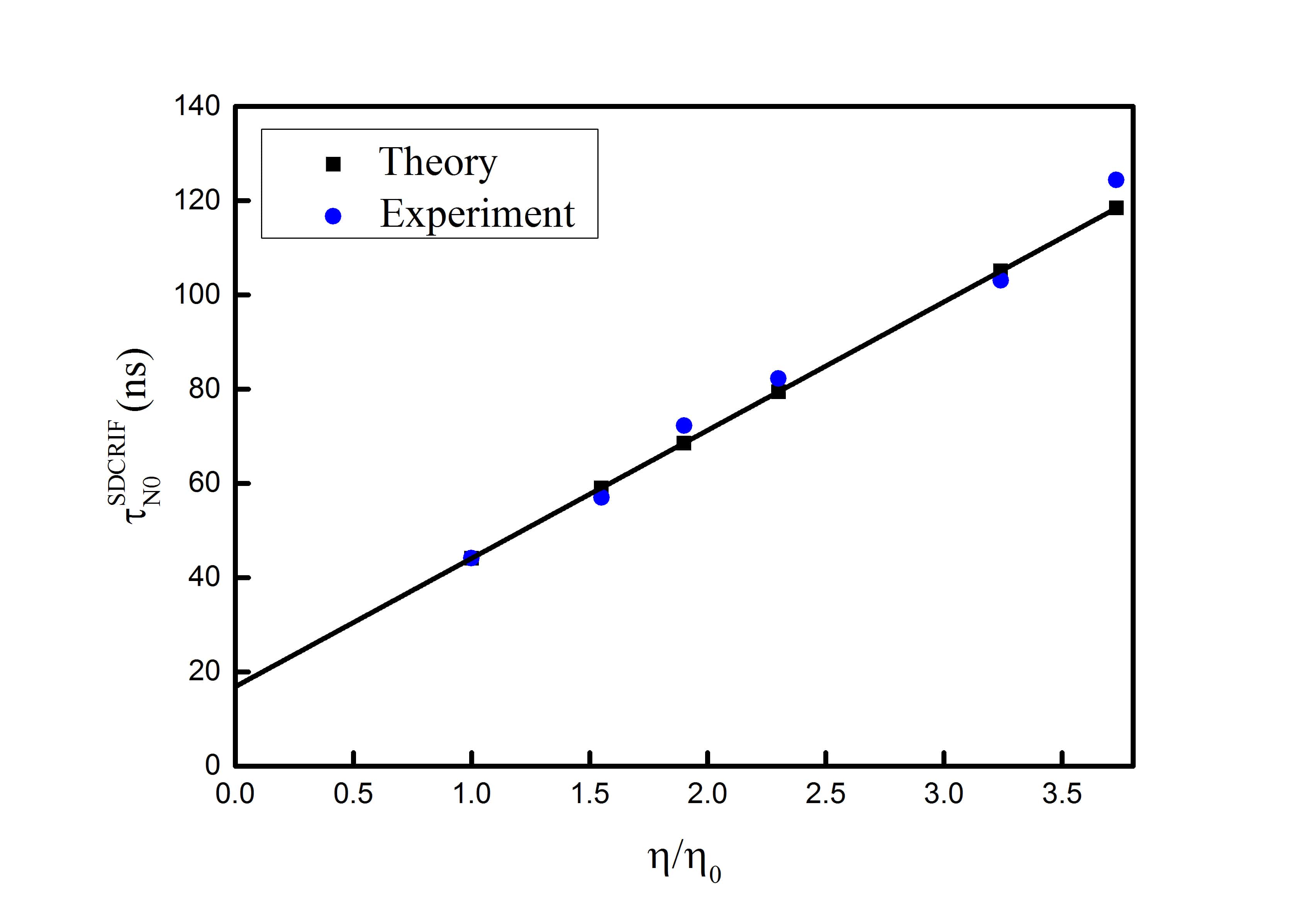}(b) \\
    \includegraphics[width=.4\textwidth]{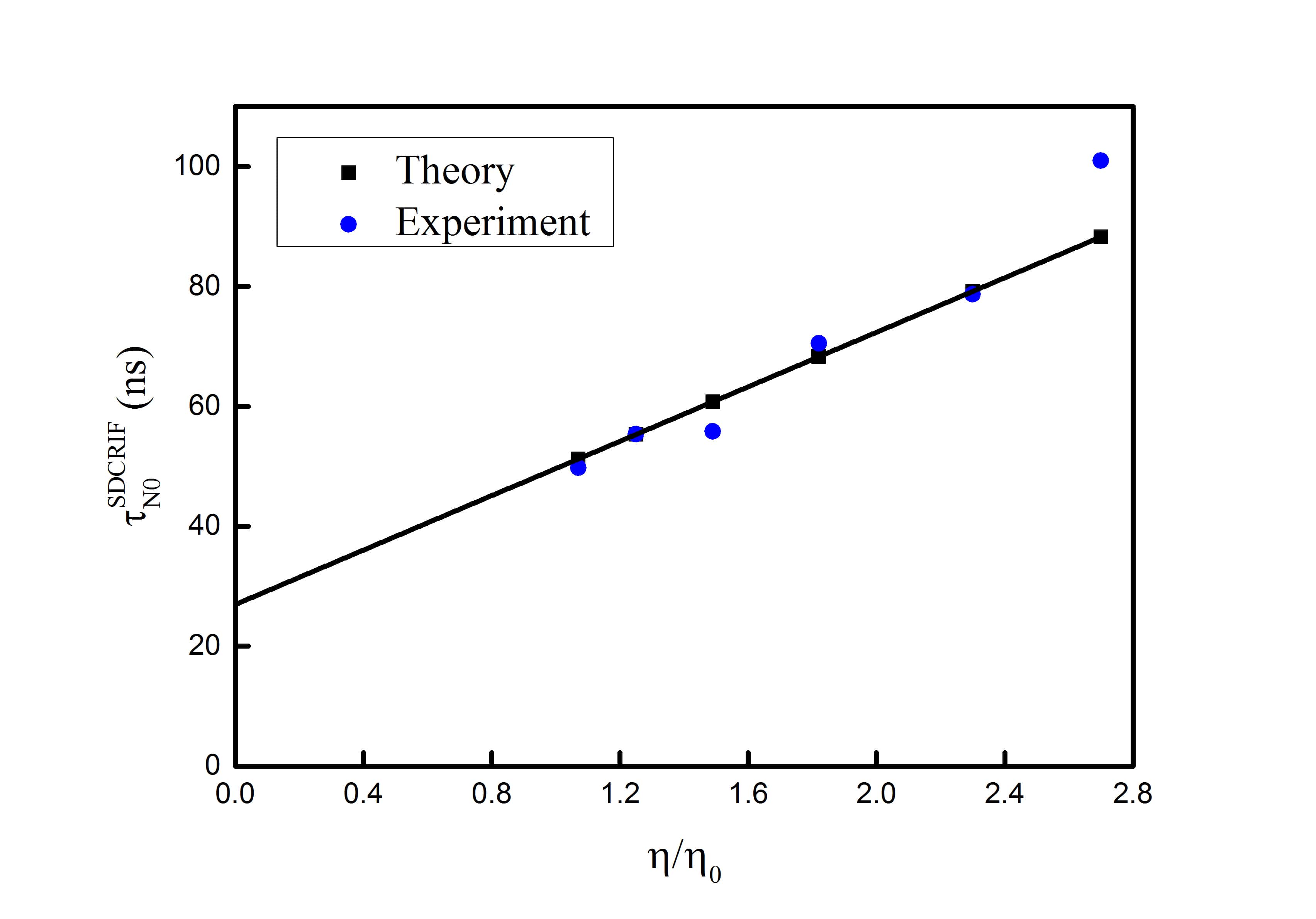}(c) &
    \includegraphics[width=.4\textwidth]{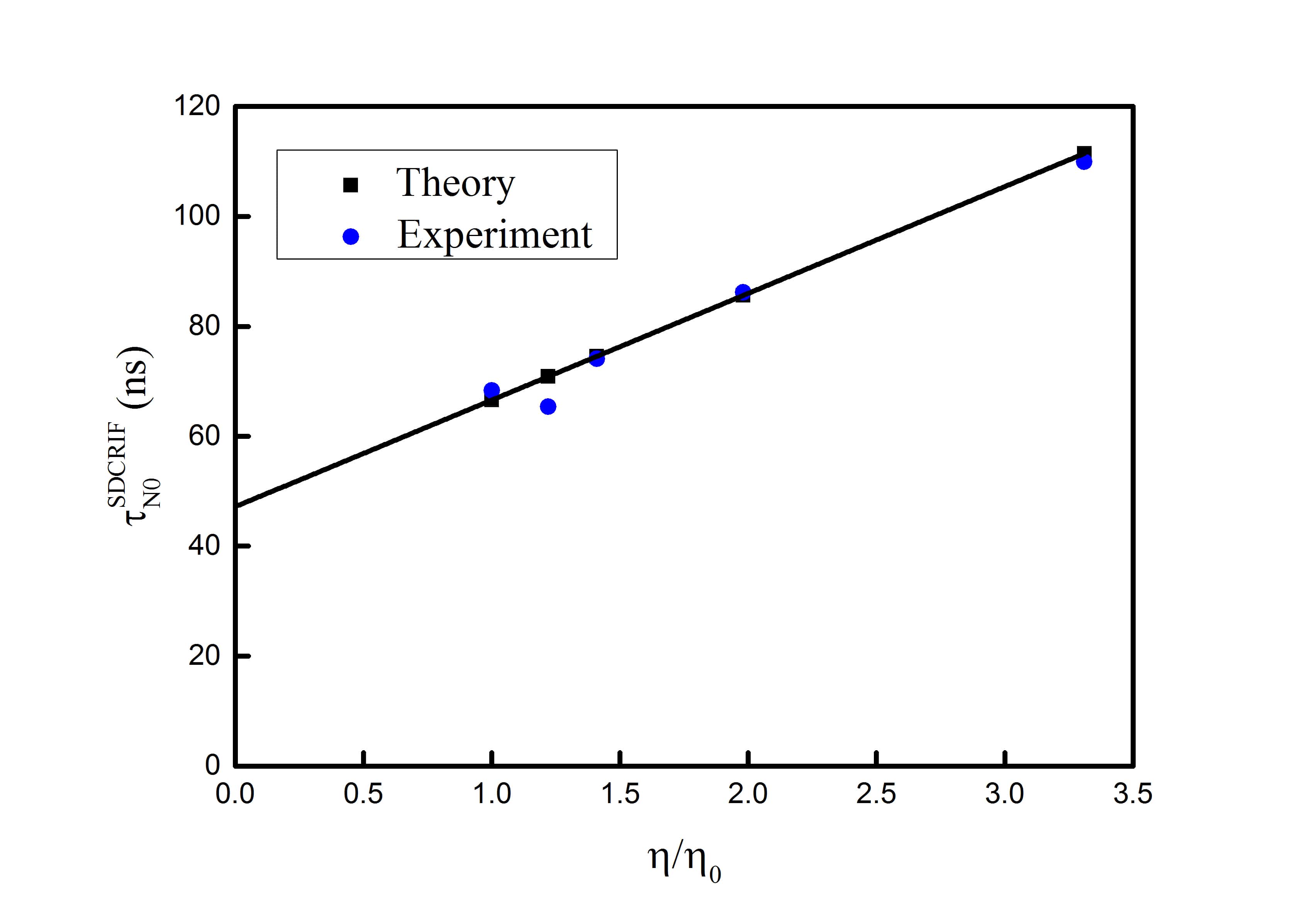}(d)
    \end{tabular}
  \caption{Comparison between theoretical and experimental reconfiguration time vs solvent viscosity data for the cold shock protein (Csp). (a) 1.3M GdmCl, (b) 2M GdmCl, (c) 4M GdmCl and (d) 6M GdmCl. The values of parameters used are $N =66$, $\tilde{k}_c=0$, $b = 3.8\times10^{-10} m$, $\xi = 9.42\times10^{-12} kgs^{-1}$, $\xi_{int,0} = 100\times\xi$, $k_B=1.38\times10^{-23} JK^{-1}$ and $T=300K$.}
  \label{fig:m}
\end{figure}

\begin{figure}
\centering
  \begin{tabular}{@{}cccc@{}}
    \includegraphics[width=.4\textwidth]{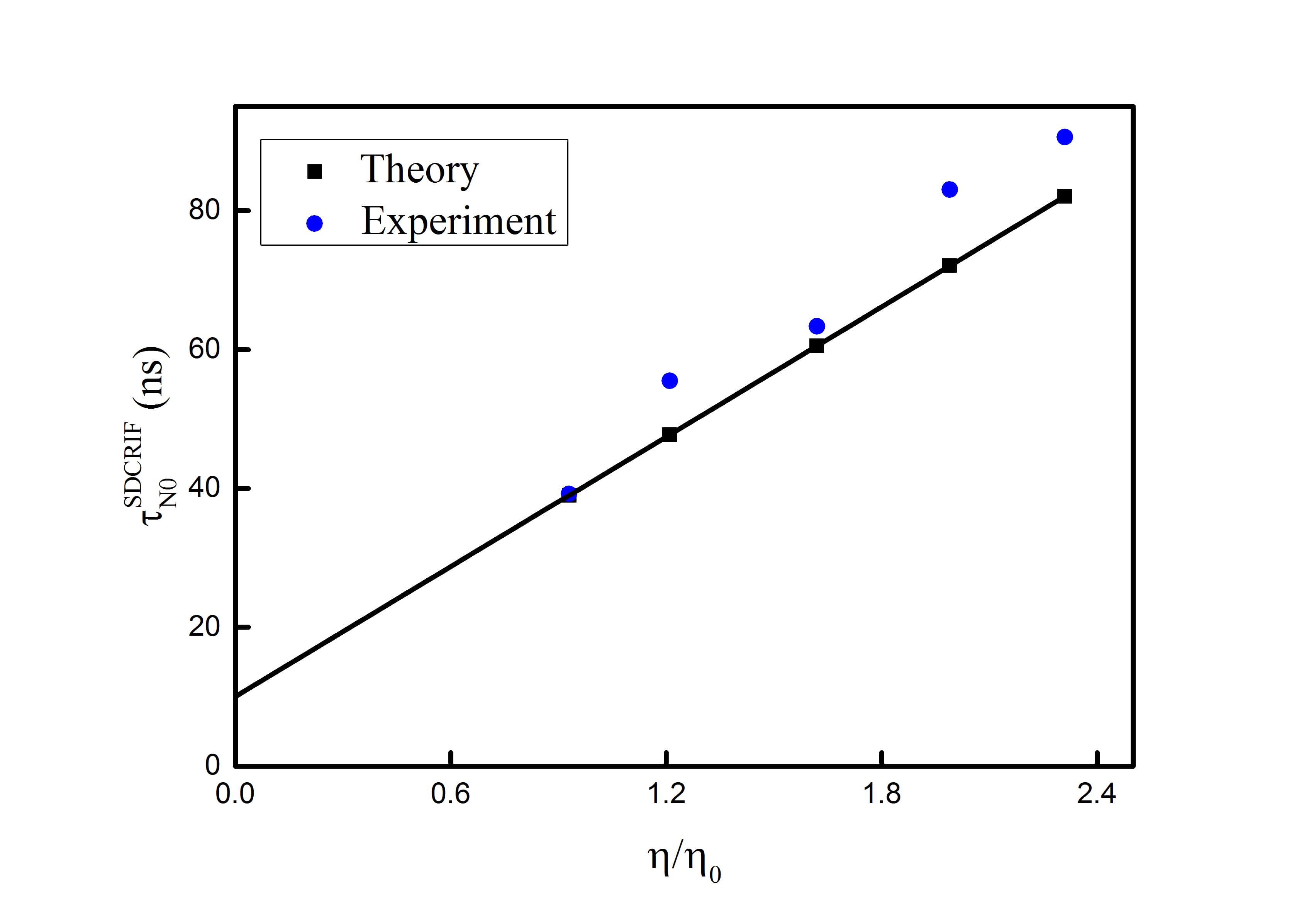} (a)
    \includegraphics[width=.4\textwidth]{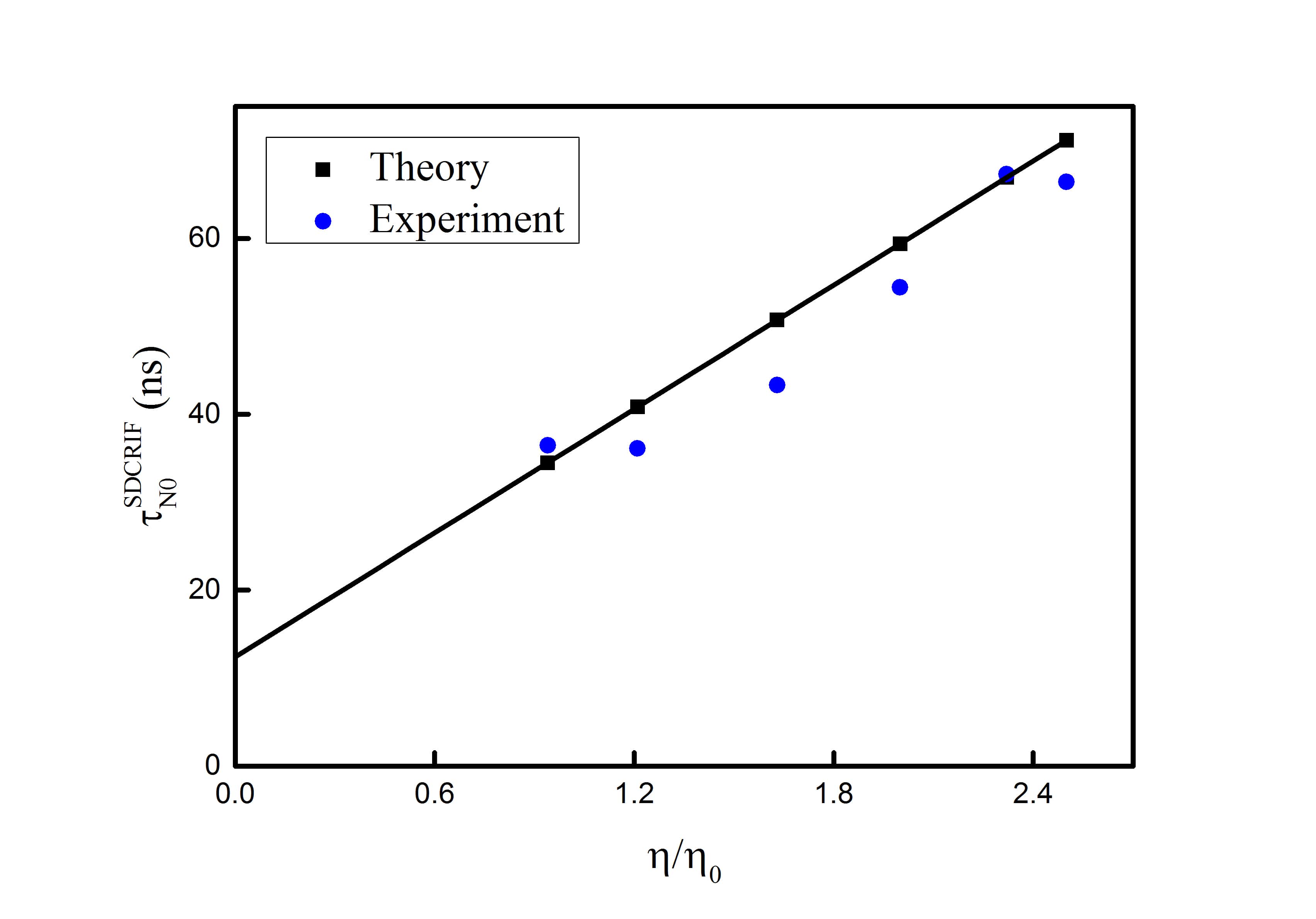} (b)& \\
    \includegraphics[width=.4\textwidth]{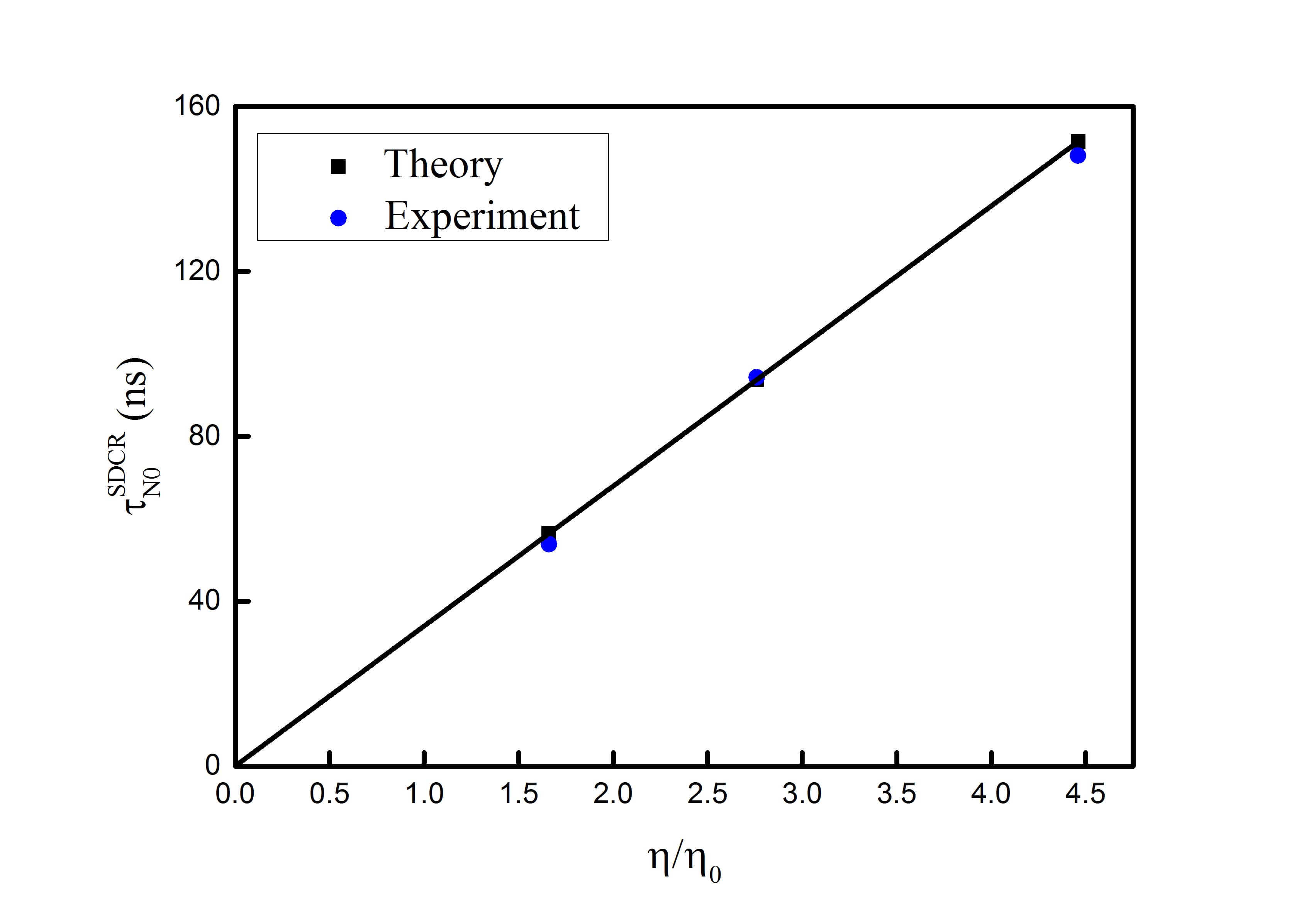} (c)
    \end{tabular}
  \caption{Comparison between theoretical and experimental reconfiguration time vs solvent viscosity data for IDP ProT$\alpha$. (a) native buffer, (b) 1M KCl and (b) 6M GdmCl. The values of parameters used are $N =110$, $b = 3.8\times10^{-10} m$, $\xi = 9.42\times10^{-12} kgs^{-1}$, $\xi_{int,0} = 100\times\xi$, $k_B=1.38\times10^{-23} JK^{-1}$ and $T=300K$.}
  \label{fig:n}
\end{figure}

In this section we use SDCRIF to compute the reconfiguration time and compare it with the one measured in the recent F\"{o}rster resonance energy transfer (FRET), nanosecond fluorescence correlation spectroscopy and microfluidic mixing based study on cold shock protein from Thermotoga maritima (Csp) labeled at positions 2 and 68 with Alexa 488 and Alexa 594 as donor and acceptor, respectively \cite{schuler2008, schuler2012}. Thus it can be approximated that the labelling are at the two ends of the protein. Their study revealed important role of internal friction in unfolded small cold shock protein and confirmed a time scale of about $\sim5-50$ ns for the internal friction. Moreover they found the internal friction to strongly depend on the denaturant concentration. Nuclear magnetic resonance and laser photolysis methods also have confirmed the effects of denaturants by showing the rate of intrachain contact formation in unfolded state of carbonmonoxide-liganded cytochrome c (cyt-CO) to increase with increase in denaturant concentration \cite{bhuyan2013}.  Higher the denaturant concentration lower the internal friction. On the other hand for an intrinsically disordered protein (IDP) such as C-terminal segment of human prothymosin $\alpha$ (ProT$\alpha$) magnitude of internal friction is smaller \cite{schuler2010, haran2010} and at high denaturant concentration it is negligibly small. This is presumably due to exposed hydrophilic and charged residues resulting expansion of the protein. But in native buffer the time scale for internal friction is $\sim6$ ns which on addition of excess salt like KCl shows $\sim3$ times increase in internal friction due to the collapse of the IDP. For both the cases the reconfiguration time measured has a linear dependence on solvent viscosity with an intercept equal to the time scale for the internal friction. Fig. (\ref{fig:m}) and Fig. (\ref{fig:n}) show the experimental data and our calculation based on SDCRIF of Csp and ProT$\alpha$ respectively. All the calculations are performed with $\nu=3/5$, value corresponding to good solvent, which is particularly valid for IDPs. On the other hand, the parameter $k_c$ in our model takes care of changes associated with the denaturant concentration and $\xi_{int}$ is the corresponding internal friction which is considered to be zero for the ProT$\alpha$ in high denaturant condition (6MGdmCl) when time scale associated with internal friction was found to be negligible experimentally. In this case $k_c=\tilde{k}_c$ and $\tilde{k}_c$ is scaled as $5.5k_{c,0}$. Table. \ref{compare} and Table. \ref{compare2} depict the parameters used in our calculation and show the theoretically calculated internal friction time scales are in excellent quantitative agreement with that measured experimentally.

 \begin{figure}
\centering
  \begin{tabular}{@{}cccc@{}}
    \includegraphics[width=.8\textwidth]{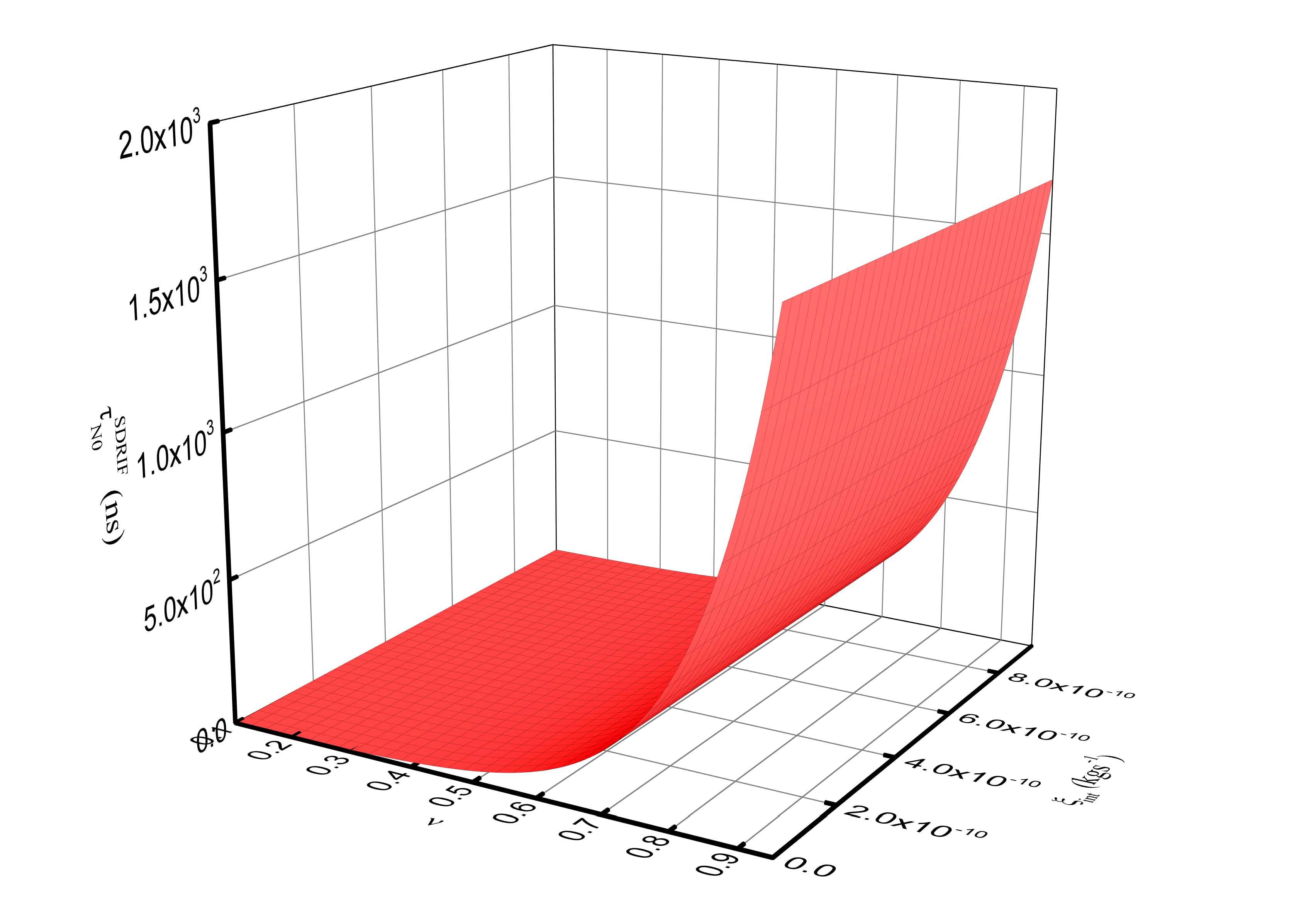} (a) \\
    \includegraphics[width=.8\textwidth]{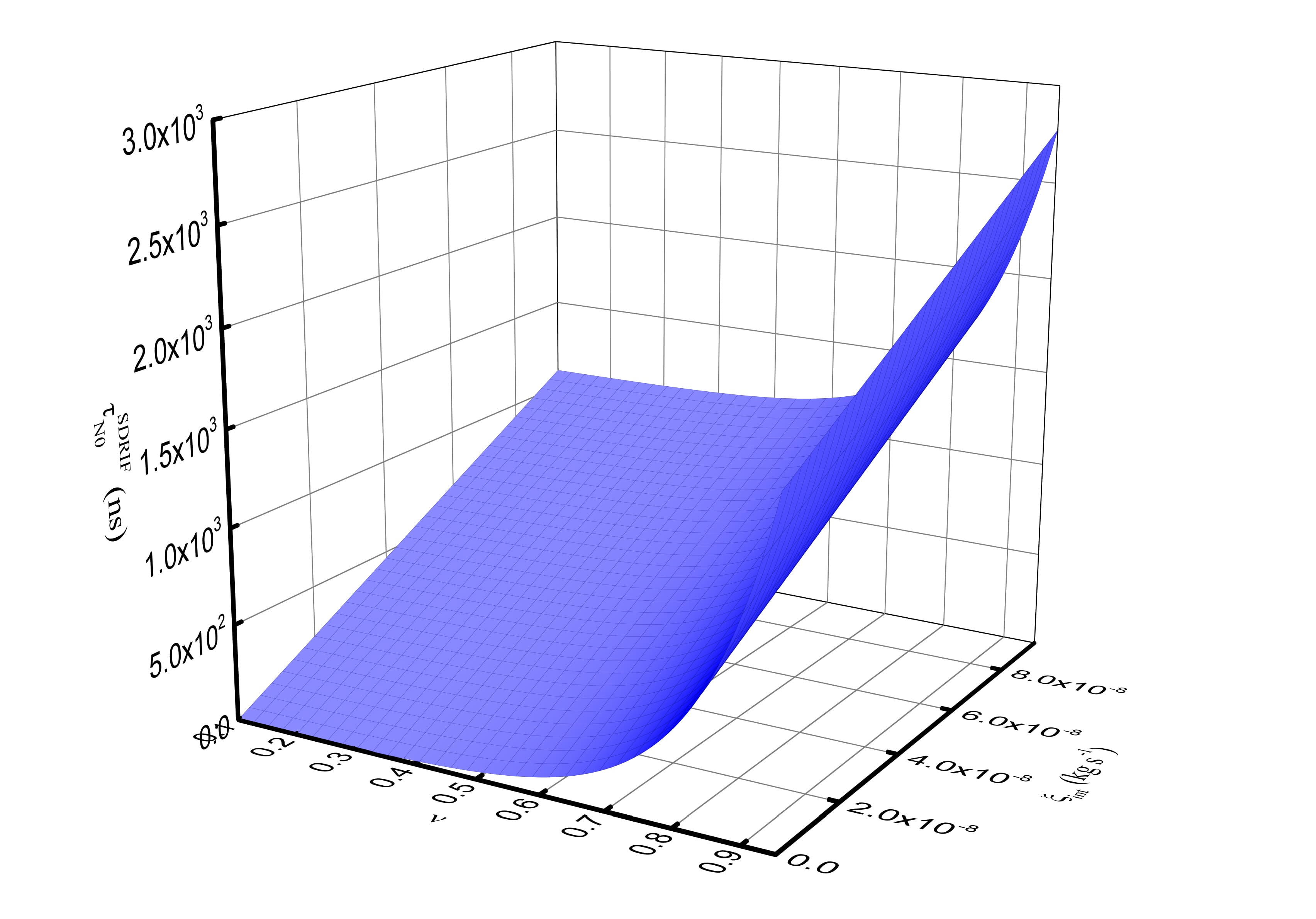} (b)
    \end{tabular}
  \caption{3D Plot of reconfiguration time vs $\nu$ and $\xi_{int}$. (a) $\xi_{int}$ upto $10^2\times \xi$, (b) $\xi_{int}$ upto $10^4\times \xi$. The values of parameters used are $N=66$, $k_c=0$, $b = 3.8\times10^{-10} m$,  $\xi = 9.42\times10^{-12} kgs^{-1}$, $k_B=1.38\times10^{-23} JK^{-1}$ and $T=300K$.}
  \label{fig:o}
\end{figure}

\section{conclusions}

Motivated by recent experiments \cite{schuler2012} on cold shock proteins and intrinsically disordered proteins (IDPs) we have analyzed the effects of denaturant and the solvent quality on the reconfiguration and looping dynamics of a chain with internal friction by using an extended Rouse chain model with internal friction. The model termed as ``Solvent Dependent Compacted Rouse Chain (SDCRIF)" takes care of solvent quality through a Flory type exponent $\nu$ and the effects of denaturant concentration are taken care by the strength of a harmonic confinement $k_c$ of the chain. Following an ansatz we further relate $k_c$ with the internal friction. This assures a non zero intercept in the plot of reconfiguration time vs solvent viscosity as found in experiments \cite{schuler2012, hofmann2012} and also makes it denaturant concentration dependent. Here we would like to point out that mere ``Compacted Rouse with Internal Friction (CRIF)" \cite{chakrabarti2015, schuler2012} can not convincingly account for the changes in reconfiguration time due to change in solvent quality and for this we need a parameter $\nu$. Also the parameter $k_c$ should be coupled strongly with the internal friction $\xi_{int}$ and this is done through the ansatz mentioned in the result section. For folded protein like cold shock protein the magnitude of the internal friction is high and our theory also reproduce this with a choice of high value of $k_c$ and $\xi_{int}$. Values of these parameters also change on changing the denaturant concentration. Typically the value of the number of monomers contributing to internal friction, $n_b$ also increases as the denaturant concentration decreases and the protein collapses which demands a higher magnitude of $k_c$ to be used to reproduce the experimental data.  On the other hand for the intrinsically disordered protein prothymosin $\alpha$ (ProT$\alpha$) magnitude of $k_c$ and the internal friction $\xi_{int}$ are low. Only in presence of a salt the protein collapses and then a higher magnitude of $k_c$ can reproduce the experimental data.  This highlights the novelty of our model which is applicable to a wide range of denaturant concentration, solvent quality and protein types. Whereas $k_c$ in our model effectively renormalizes the force constant of the chain, $\nu$ on the other hand takes care of the quality of the solvent around. Both the parameters are essential to explain the experimental data on reconfiguration times of proteins, so as the azsatz that relates $k_c$ with the internal friction $\xi_{int}$ of the chain. Another issue is the relative dependence of the reconfiguration and looping time on $\xi_{int}$ and $\nu$. For the range of values of $\xi_{int}$ we have used throughout our calculation the reconfiguration time (so as the looping time, not shown) depends only weakly on internal friction $\xi_{int}$ and rather strongly on the solvent quality, i.e. $\nu$. This can be seen in Fig. (\ref{fig:o}) (where $c(\nu)$ is taken to be $0.80$ for simplicity) but on increasing the internal friction magnitude by two orders of magnitude the change due internal friction become visible. For example if the $\xi_{int}$ is taken to be $3000\xi$ for poor solvent $(\nu=1/3)$ the time scale for internal friction become ($\sim340$ ns) which is $\sim3$ times higher than the same ($\sim100$ ns) in good solvent $(\nu=3/5)$ for $\xi_{int}=100\xi$. This is in the same spirit as seen in recent simulation \cite{luo2015}, where the looping time passes through a minima while plotted against the parameter $\lambda$, a combined measure of solvent quality and internal friction.

We would like to conclude by pointing out that in reality internal friction has contributions from  hydrogen bonding, other weak forces and specially torsion angle rotations in proteins and have already been investigated in atomistic simulations \cite{best2014, best2015, makarov2014}. Taking these contributions explicitly beyond the scope of this study.  However it would be worth incorporating internal friction in a model of polymer with torsion and semiflexibility \cite{kawakatsubook} and explore the physics involved. Work along this direction is under progress.

\section{acknowledgement}

Authors thank IRCC IIT Bombay (Project Code: 12IRCCSG046), DST (Project No. SB/SI/PC-55/2013) and CSIR (Project No. 01(2781)/14/EMR-II) for funding generously.

\begin{table}[tbp]
\begin{tabular}{|c||c||c||c||c|}
\hline Denaturant concentration & $k_c$ & $\xi_{int}$ & $\tau_{int}$(Theory) & $\tau_{int}$(Experiment)   \\ \hline
$1.3$M GdmCl & $3.0k_{c,0}$ & $9.0\xi_{int,0}$& $\sim47$ ns &  $\sim42$ ns \\ \hline
$2.0$M GdmCl& $2.5k_{c,0}$ &$5.0\xi_{int,0}$ &$\sim27$ ns& $\sim25$ ns \\
\hline
$4.0$M GdmCl& $2.0k_{c,0}$ & $3.0\xi_{int,0}$& $\sim17$ ns& $\sim12$ ns \\
\hline
$6.0$M GdmCl&$1.5k_{c,0}$  & $1.5\xi_{int,0}$ & $\sim9$ ns& $\sim5$ ns\\ \hline
\end{tabular}
\caption{Comparison with experimental data on cold shock protein: The values of parameters used are $N =66$, $\tilde{k}_c=0$, $b = 3.8\times10^{-10} m$, $\xi = 9.42\times10^{-12} kgs^{-1}$, $\xi_{int,0} = 100\times\xi$, $k_B=1.38\times10^{-23} JK^{-1}$ and $T=300K$. }
\label{compare}
\end{table}

\begin{table}[tbp]
\begin{tabular}{|c||c||c||c||c|}
\hline Solvent condition & $k_c$ & $\xi_{int}$ &$\tau_{int}$(Theory) & $\tau_{int}$(Experiment)  \\
\hline
Native buffer&$\tilde{k}_c=0,6k_{c,0}$ & $2.0\xi_{int,0}$& $\sim10$ ns &  $\sim6$ ns \\ \hline
$1$M KCl&$\tilde{k}_c=0,8k_{c,0}$ &$2.5\xi_{int,0}$ &$\sim12$ ns& $\sim16$ ns \\
\hline
$6$M GdmCl & $\tilde{k}_c\neq0,k_{c,0}=0$ & $0$& $0$& $0$ \\
\hline

\end{tabular}
\caption{Comparison with experimental data on IDP ProT$\alpha$: The values of parameters used are $N =110$, $b = 3.8\times10^{-10} m$, $\xi = 9.42\times10^{-12} kgs^{-1}$, $\xi_{int,0} = 100\times\xi$, $k_B=1.38\times10^{-23} JK^{-1}$ and $T=300K$.}
\label{compare2}
\end{table}

%\bibliography
%\bibliographystyle{apsrev}
%\bibliography{ringclosurenew}
%\end{bibliography}

\end{document}